\documentclass[main]{aa}
\usepackage{graphicx}
\usepackage{txfonts}

\def\Lsun{L$_\odot$}
\def\Msun{M$_\odot$}

\def\Oiii{[O\,{\sc iii}]}

\def\Oii{[O\,{\sc ii}]}
\def\Oi{[O\,{\sc i}]}
\def\Nii{[N\,{\sc ii}]}

\def\Cii{[C\,{\sc ii}]}
\def\Ci{[C\,{\sc i}]}

\def\kms{km\,s$^{-1}$}

\def\Kkmspc{K~km\,s$^{-1}$\,pc$^2$}
\def\lsim{\mathrel{\rlap{\lower 3pt \hbox{$\sim$}} \raise 2.0pt \hbox{$<$}}}
\def\gsim{\mathrel{\rlap{\lower 3pt \hbox{$\sim$}} \raise 2.0pt \hbox{$>$}}}

\begin{document}

\authorrunning{Decarli et al.}
\titlerunning{ISM in a $z\sim 6.4$ quasar host.}

\title{A comprehensive view of the interstellar medium in a quasar host galaxy at $z\approx 6.4$}
\author{
Roberto Decarli\inst{1}, 
Antonio Pensabene\inst{2},
Tanio Diaz-Santos\inst{3,4},
Carl Ferkinhoff\inst{5},
Michael A.~Strauss\inst{6},
Bram P.~Venemans\inst{7},
Fabian Walter\inst{8},
Eduardo Ba\~{n}ados\inst{8},
Frank Bertoldi\inst{9},
Xiaohui Fan\inst{10},
Emanuele Paolo Farina\inst{11},
Dominik A.~Riechers\inst{12},
Hans-Walter Rix\inst{8},
Ran Wang\inst{13}
}
\institute{
INAF -- Osservatorio di Astrofisica e Scienza dello Spazio di Bologna, via Gobetti 93/3, I-40129, Bologna, Italy. \email{ roberto.decarli@inaf.it} \and
Dipartimento di Fisica ``G. Occhialini'', Universit\`a degli Studi di Milano-Bicocca, Piazza della Scienza 3, I-20126, Milano, Italy. \and
Institute of Astrophysics, Foundation for Research and Technology-Hellas (FORTH), Heraklion, 70013, Greece. \and
School of Sciences, European University Cyprus, Diogenes street, Engomi, 1516 Nicosia, Cyprus. \and
Department of Physics, Winona State University, Winona, MN 55987, USA. \and
Department of Astrophysical Sciences, Princeton University, 4 Ivy Lane, Princeton, NJ 08544, USA.\and
Leiden Observatory, Leiden University, P.O. Box 9513, NL-2300 RA Leiden, The Netherlands. \and
Max-Planck Institut f\"{u}r Astronomie, K\"{o}nigstuhl 17, D-69117, Heidelberg, Germany. \and
Argelander-Institute for Astronomy, University of Bonn, Auf dem H\"{u}gel 71, D-53121 Bonn, Germany. \and
Steward Observatory, University of Arizona, 933 N. Cherry St., Tucson, AZ  85721, USA. \and
Gemini Observatory, NSF's NOIRLab, 670 N A'ohoku Place, Hilo, Hawai'i 96720, USA. \and
I.~Physikalisches Institut, Universit\"{a}t zu K\"{o}ln, Z\"{u}lpicher Strasse 77, 50937 K\"{o}ln, Germany.\and
Department of Astronomy, School of Physics, Peking University, 5 Yiheyuan Road, Haidian District, Beijing 10087, PR China.
}

\date{February 2023}
\abstract{
Characterizing the physical conditions (density, temperature, ionization state, metallicity, etc) of the interstellar medium is critical to our understanding of the formation and evolution of galaxies. Here we present a multi-line study of the interstellar medium in the host galaxy of a quasar at $z\approx6.4$, i.e., when the universe was 840 Myr old. This galaxy is one of the most active and massive objects emerging from the dark ages, and therefore represents a benchmark for models of the early formation of massive galaxies. We used the Atacama Large Millimeter Array to target an ensemble of tracers of ionized, neutral, and molecular gas, namely the fine--structure lines: \Oiii{} 88\,$\mu$m, \Nii{} 122\,$\mu$m, \Cii{} 158\,$\mu$m, and \Ci{} 370\,$\mu$m and the rotational transitions of CO(7--6), CO(15-14), CO(16-15), and CO(19-18); OH 163.1\,$\mu$m and 163.4\,$\mu$m; and H$_2$O 3(0,3)-2(1,2), 3(3,1)-4(0,4), 3(3,1)-3(2,2), 4(0,4)-3(1,3), 4(3,2)-4(2,3). All the targeted fine-structure lines are detected, as are half of the targeted molecular transitions. By combining the associated line luminosities, the constraints on the dust temperature from the underlying continuum emission, and predictions from photoionization models of the interstellar medium, we find that the ionized phase accounts for about one third of the total gaseous mass budget, and is responsible for half of the total \Cii{} emission. It is characterized by high density ($n\sim180$\,cm$^{-3}$), typical of HII regions. The spectral energy distribution of the photoionizing radiation is comparable to that emitted by B-type stars. Star formation also appears to drive the excitation of the molecular medium. We find marginal evidence for outflow-related shocks in the dense molecular phase, but not in other gas phases. This study showcases the power of multi-line investigations in unveiling the properties of the star-forming medium in galaxies at cosmic dawn.
}
\maketitle

\section{Introduction} 

Quasar host galaxies are among the first massive galaxies emerging from the dark ages ($z\gsim6$). They host black holes with masses that can exceed $10^9$\,\Msun{} \citep[e.g.,][]{willott03,derosa11,wu15,banados18a}. They often form stars at high rates \citep[SFR=100--1,000\,\Msun{}\,yr$^{-1}$;][]{bertoldi03a,walter04,walter09a,wang08a,wang08b,wang13,leipski14,drouart14,venemans18,decarli18}. Immense gaseous reservoirs ($M_{\rm H2}\gsim 10^{10}$\,\Msun{}) fuel this intense nuclear and star formation activity  \citep[e.g.,][]{walter03,bertoldi03b,wang08a,venemans17a}. The luminosity of emission lines associated with heavy elements and ions in the broad-line region \citep{kurk07,derosa11,derosa14,schindler20}, as well as the presence of large reservoirs of dust in the interstellar medium \citep{wang08a,venemans18} indicate that these galaxies are already highly metal--enriched. The mass, star formation rate, and metallicity of these early quasar host galaxies exceed by orders of magnitudes those of typical star--forming galaxies at $z>6$ \citep[see, e.g.,][]{vanzella14,bouwens15,oesch16,harinake17,banados18b,ota18,salmon18,curti22}. In this respect, quasar host galaxies are arguably the most active and some of the most evolved objects emerging from the dark ages. They represent a challenge for models of early black hole and galaxy coexistence \citep{volonteri12,habouzit16,habouzit17,agarwal17,yue17,lupi19,lupi22,romano23}. Characterizing their early growth is therefore a mandatory step to understand how first massive galaxies formed. 

Direct observations of the starlight in quasar host galaxies at $z\gsim6$ are hindered by the unfavorable contrast against the bright nuclear emission from the quasar, and by a combination of redshift and dust obscuration that suppresses the observed rest-frame UV emission and shifts the rest-frame optical emission into the mid-infrared bands, where the atmosphere is opaque. These limitations also affect nebular gas tracers (the hydrogen Balmer lines: H$\alpha$, H$\beta$, H$\gamma$, etc; ionized and neutral oxygen: \Oiii{} at 5008\,\AA{}, \Oii{} at 3727\,\AA{}, \Oi{} at 6300\,\AA{}; ionized nitrogen \Nii{} at 6584\,\AA{}; etc) and will only be effectively overcome by the {\em James Webb Space Telescope}. On the other hand, a wealth of information on the star--forming medium at $z\gsim6$ is accessible nowadays at sub-mm and mm wavelengths. Rest-frame far--infrared (IR) gas tracers detected in distant quasar host galaxies include various rotational molecular transitions from, e.g., carbon monoxide, CO; water, H$_2$O; hydroxyl, OH \citep[see, e.g.,][]{walter03,weiss07,wang11,vanderwerf11,omont11,omont13,riechers13,riechers14,venemans17a,wang19,yang19,li20,pensabene21,decarli22}. In addition, many key elements and ions have fine-structure transitions at these wavelengths, such as ionized carbon \Cii{} at 158 $\mu$m; neutral carbon \Ci{} at 370\,$\mu$m and 609\,$\mu$m; ionized nitrogen \Nii{} at 122\,$\mu$m and 205\,$\mu$m; ionized oxygen \Oiii{} at 52 and 88\,$\mu$m; and neutral oxygen \Oi{} at 63\,$\mu$m and 146\,$\mu$m. Many of these lines have been detected in high--redshift galaxies over the last few years \citep[e.g.,][]{maiolino05,maiolino09,walter09a,walter11,walter12,ferkinhoff11,ferkinhoff15,coppin12,combes12,nagao12,decarli12,decarli14,venemans12,venemans17a,venemans17b,carilli13,brisbin15,gullberg15,capak15,pavesi16,uzgil16,bothwell17,trakhtenbrot17,willott17,lamarche17,lamarche19,carniani18,hashimoto19,novak19,tadaki19,boogaard20,valentino20,sugahara21,harrington21,meyer22}. Combinations of these emission lines can shed light on the gas mass in various phases of the interstellar medium (ISM; see, e.g., \citealt{ferkinhoff15,zanella18,dunne21} and the review of \citealt{carilli13}); on the star formation rate (SFR; see, e.g., \citealt{delooze14,herreracamus15,herreracamus18}); on metallicity ($Z$; see, e.g., \citealt{nagao12,peng21,lamarche22}); on the gas density ($n$), on the origin of the excitation mechanism, and on the strength and hardness of the incident flux \citep[for a recent review, see][]{wolfire22}. While mapping individual clouds in multiple ISM tracers at high redshift is impossible even with the excellent sensitivity and angular resolution provided by the Atacama Large Millimeter Array (ALMA), galaxy--averaged observations of these tracers can shed light on global properties of the ISM (e.g., on the mass budget in the ionized, neutral, and molecular phases; on the contribution of the central quasar to the ISM excitation; and on the total metallicity of the galaxy; see, e.g., \citealt{romano23}).

In this work, we present a multi-line study of the star--forming medium in the host galaxy of the quasar PSO J183.1124+05.0926 (hereafter, PJ183+05; R.A.: 12:12:26.97, Dec.: +05:05:33.4) at $z=6.4386$. This quasar was discovered by \citet{banados16} using color-color selections from the Pan-STARRS1 database \citep{chambers16}, and follow-up photometric and spectroscopic observations. The \Cii{} luminosity in this source is the highest among 27 quasars at $z>6$ surveyed in \citet{decarli18}, and one of the highest from non-lensed sources currently known at any redshift \citep[e.g.,][]{pavesi16,tadaki19,andika20,mitsuhashi21}. The quasar resides at a redshift that is convenient for observations of various far-infrared emission lines that fall in transparent windows of the atmosphere at $z\approx 6.4$. We targeted a suite of lines sampling the molecular, neutral and ionized components of the ISM in PJ183+05. The structure of the paper is as follows: In Sec.~\ref{sec_observations} we present the observations and the data processing. In Sec.~\ref{sec_analysis}, we analyze the dataset. In Sec.~\ref{sec_results} we infer physical quantities from the observed spectra, and discuss our findings. Finally, we draw our conclusions in Sec.~\ref{sec_conclusions}.

Throughout the paper we adopt a flat $\Lambda$CDM cosmology with $H_0=70$ km\,s$^{-1}$\,Mpc$^{-1}$, $\Omega_{\rm m}=0.3$ and $\Omega_{\Lambda}=0.7$ \citep[consistent with the measurements by the][]{planck15}. Within this cosmological framework, $z=6.4386$ corresponds to a luminosity distance of $D_{\rm L}=62671$\,Mpc and an angular scale of 5.491 kpc per arcsec.

\section{Observations and data reduction}\label{sec_observations}

The dataset used in this project consists of the \Cii{} 158\,$\mu$m observations of PJ183+05 presented in \citet{decarli18} (program ID: 2015.1.01115.S, PI: Walter); as well as data from a dedicated ALMA program (ID: 2016.1.00226.S, PI: Decarli). Observations were carried out in compact array configurations. Table\,\ref{tab_obs} summarizes the observations. Integration times ranged between 8 and 80 min in each frequency setting. The quasars J1229+0203 and J1222+0413 served as bandpass / pointing and phase calibrators, respectively. Ganymede was observed for flux calibration, with the exception of the band 8 observations for which the quasar J1229+0203 served as flux calibrator.

We targeted the following fine--structure lines: \Cii{} at 158\,$\mu$m, \Nii{} 122\,$\mu$m, \Ci{} 370\,$\mu$m, and \Oiii{} 88\,$\mu$m. 
The \Oi{} 63\,$\mu$m line was also scheduled for observations, but unfavorable weather conditions prevented the completion of the program within Cycle 4. In addition, we targeted four rotational lines from carbon monoxide: CO(7--6), CO(15--14), CO(16--15), and CO(19--18); the OH doublet at 163.1 and 163.4 $\mu$m; and five water vapor transitions: H$_2$O 4(3,2)--4(2,3), 4(0,4)--3(1,3), 3(3,1)--3(2,2), 3(3,1)--4(0,4), 3(0,3)--2(1,2)\footnote{Technically, we also have coverage for other water transitions at very high J, such as the 8(4,5)--8(3,6) transition at 3495.358 GHz. However, these transitions are expected to be extremely weak, and indeed none of the J$>$4 lines is detected. We thus restrict our analysis to J$\leq$4 transitions.}. 

Data were reduced and calibrated with the official ALMA pipeline in CASA \citep[][version 4.7.2]{mcmullin07}. We imaged the measurement sets using natural weighting, in order to maximize the signal-to-noise ratio of line detections. The \Cii{} data presented in \citet{decarli18} was binned in 30\,\kms{} channels. The other line observations presented here are binned in 90\,\kms{} wide channels. The expected line width ($\sim 375$\,\kms{} from the analysis of the \Cii{} data) is thus sampled in $\sim4$ independent channels, while at the same time maximizing the signal-to-noise ratio in the lines. From each frequency setting, we create two cubes, corresponding to the lower and upper side bands. The cubes are cleaned to the 2-$\sigma$ level, using cleaning masks on the quasar. 

\begin{figure*}
\begin{center}
\includegraphics[width=0.33\textwidth]{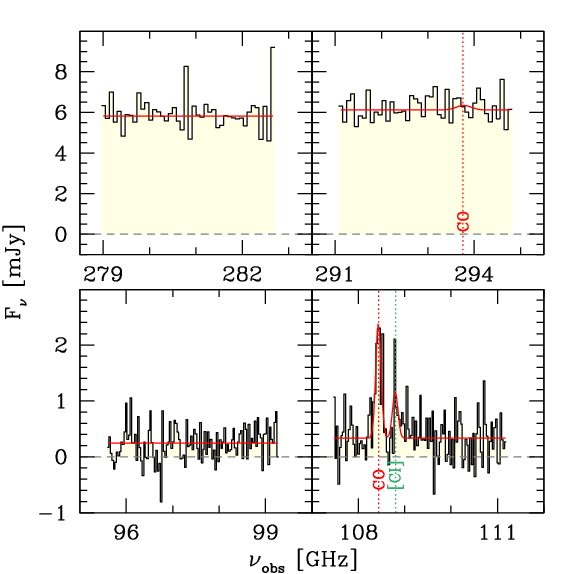}
\includegraphics[width=0.33\textwidth]{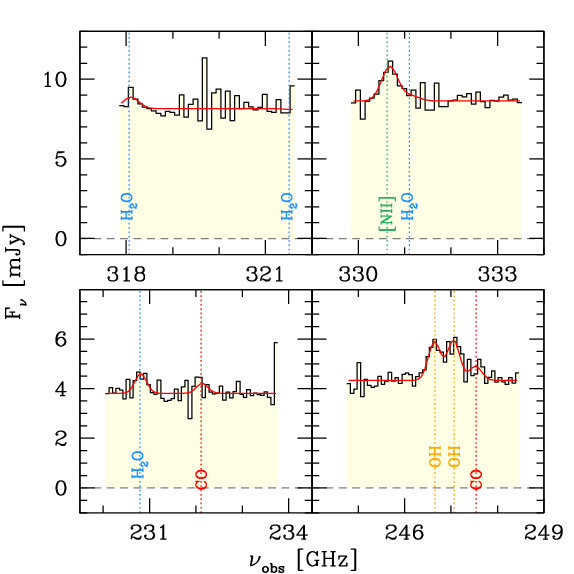}
\includegraphics[width=0.33\textwidth]{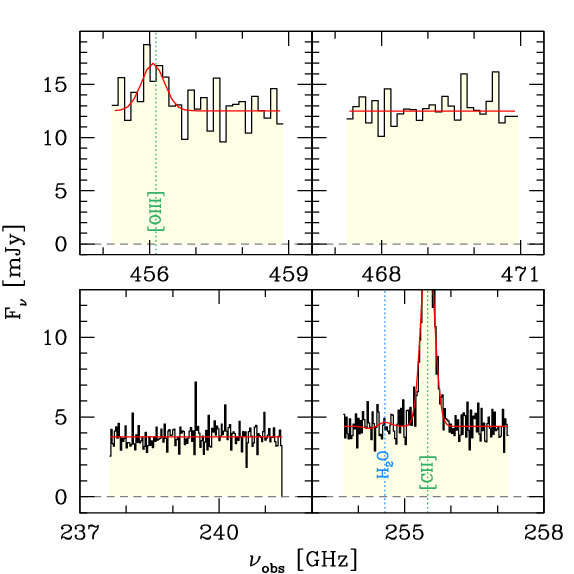}\\
\end{center}
\caption{Observed ALMA spectra of PJ183+05 (histograms). The fitted continuum+line models are shown as thick red lines. Main transitions are marked with vertical dotted lines.}
\label{fig_spec}
\end{figure*}

Two-dimensional channel-by-channel fits of the quasar emission at 88\,$\mu$m in the rest-frame revealed that the emission is spatially unresolved or only marginally resolved at $>0.7''$ resolution \citep[see also the analysis of the \Cii{} emission presented in][]{venemans20}. Therefore, we extracted the spectra of the dust continuum and targeted lines in PJ183+05 on a single-pixel basis. The extracted spectra are shown in Fig.~\ref{fig_spec}.

\begin{table*}
\caption{Summary of the observations. The RMS refers to the observed noise per channel in the cleaned data cube, for a channel width of 90\,\kms{}, except for \Cii{} (setup 4) which is computed over channels of 30\,\kms{} (see text).}\label{tab_obs}
\vspace{-5mm}
\begin{center}
\begin{tabular}{c|cccccc}
\hline
Setup                      & 1                  & 2		     & 3		  & 4		       & 5		    & 6 		 \\	
\hline
Obs.Date                   & 2017-03-27 	& 2017-03-22	     & 2017-03-19	  & 2016-01-27         & 2017-03-19	    & 2017-01-04	 \\	
Band                       &  8 		& 7		     & 7		  & 6		       & 6		    & 3 		 \\	
Ref.frequency [GHz]        & $456.12$		& $330.79$	     & $293.89$ 	  & $255.50$	       & $245.69$	    & $108.40$  	 \\	
Beam                       & $0.9''\times0.8''$ & $1.3''\times1.1''$ & $1.8''\times1.0''$ & $0.8''\times0.7''$ & $1.6''\times1.3''$ & $4.0''\times3.3''$ \\	
Int.Time           [s]     & 4890		& 3878  	     & 3126		  &  514	       & 2141		    & 2005		 \\	
RMS [$\mu$Jy\,beam$^{-1}$] & $415$		& $337$	             & $414$		  & $503$	       & $257$		    & $345$		 \\	
\hline
\end{tabular}
\end{center}
\end{table*}

\begin{table*}
\caption{Line measurements. Quoted limits are at 3-$\sigma$ significance. The last column lists the opacity corrections computed for the best fit of the dust continuum described in sec.~\ref{sec_continuum}. }\label{tab_lines}
\vspace{-5mm}
\begin{center}
\begin{tabular}{cc|cccc}
\hline
Line 		         & $\nu_0$ & $z$              & $F_\nu^{\rm line}$ & $L_{\rm line}$     & $e^{-\tau_\nu}$ \\
     		         & [GHz]   &   		      & [Jy\,\kms]         & [$10^8$\,\Lsun]    &         \\
\hline
\multicolumn{6}{c}{\it Fine structure lines} 						                  \\
\hline
\Oiii$_{88\,\mu{\rm m}}$ & 3393.01 & $6.4396\pm0.0002$ & $1.79\pm0.12	      $ &  $33.3\pm2.3$ &  $0.22$ \\   
\Nii$_{122\,\mu{\rm m}}$ & 2459.38 & $6.4373\pm0.0002$ & $0.87\pm0.08	      $ &  $11.8\pm1.1$ &  $0.45$ \\   
\Cii$_{158\,\mu{\rm m}}$ & 1900.55 & $6.4386\pm0.0001$ & $5.00\pm0.09	      $ &  $52.2\pm0.9$ &  $0.63$ \\   
\Ci$_{370\,\mu{\rm m}}$  &  809.34 & $6.4391\pm0.0005$ & $0.33\pm0.11	      $ &  $1.5\pm0.5$  &  $0.92$ \\ 	
\hline												    	
\multicolumn{6}{c}{\it Carbon monoxide}							 	    	  \\
\hline												    	
CO(7--6)		 &  806.65 & $6.4391\pm0.0005$ & $0.82\pm0.09	      $ &  $3.6\pm0.4$  &  $0.92$ \\   
CO(15--14)		 & 1726.60 & ---	       & $<0.25 	      $ &  $<2.3$	&  $0.68$ \\   
CO(16--15)		 & 1841.34 & ---	       & $<0.27 	      $ &  $<2.7$	&  $0.64$ \\   
CO(19--18)               & 2185.13 & ---	       & $<0.27 	      $ &  $<3.3$	&  $0.53$ \\   
\hline												    	
\multicolumn{6}{c}{\it Hydroxyl}								    	  \\
\hline												    	
OH$_{163.1\,\mu{\rm m}}$ & 1837.82 & $6.4389\pm0.0001$ & $0.63\pm0.07	      $ &  $6.3\pm0.7$  &  $0.65$ \\   
OH$_{163.4\,\mu{\rm m}}$ & 1834.75 & $6.4389\pm0.0001$ & $0.65\pm0.07	      $ &  $6.5\pm0.8$  &  $0.65$ \\   
\hline												    	     
\multicolumn{6}{c}{\it Water vapor}								    	  \\ 
\hline												    	     
H$_2$O 4(3,2)--4(2,3)    & 2462.93 & ---	       & $<0.27 	      $ &  $<3.7$	&  $0.45$ \\   
H$_2$O 4(0,4)--3(1,3)    & 2391.57 & ---	       & $<0.30 	      $ &  $<3.9$	&  $0.47$ \\   
H$_2$O 3(3,1)--3(2,2)    & 2365.90 & $6.4376\pm0.0003$ & $0.29\pm0.08	      $ &  $3.7\pm1.0$  &  $0.48$ \\   
H$_2$O 3(3,1)--4(0,4)    & 1893.69 & ---	       & $<0.30 	      $ &  $<3.1$	&  $0.63$ \\   
H$_2$O 3(0,3)--2(1,2)    & 1716.77 & $6.4380\pm0.0004$ & $0.32_{-0.08}^{+0.04}$ &  $3.0\pm0.7$  &  $0.69$ \\   
\hline
\end{tabular}
\end{center}
\end{table*}

\section{Analysis}\label{sec_analysis}

\subsection{Spectral fits}\label{sec_data}

We model the observed spectra as a flat continuum plus a Gaussian profile for each line. This is shown to provide a good description of the typical line profiles observed in $z>6$ quasars \citep{decarli18}. For the fitting, we used our custom Monte Carlo Markov Chain tool to sample the posterior probability of each fitted parameter. Because of the limited S/N of some of the lines in our new data, we opt to assume a fixed Gaussian profile for each transition, with a line full width at half maximum of 375\,\kms{} as observed for the \Cii{} line \citep{decarli18}. The redshift of the line, its normalization, and the underlying continuum flux density are left free in the fit. As priors, we used a Gaussian with a standard deviation of 0.002 around the \Cii{} redshift, corresponding to $\sim 80$\,\kms{}; a Gaussian distribution for the line normalization (centered on a rough estimate of the line flux based on the line peak and assumed width); and a Maxwellian distribution for the continuum flux density, with a scale width based on the median value of the spectrum in each frequency tuning.

Figure~\ref{fig_spec} shows the spectra and their fits for all the frequency settings used in this work. We list the results of the fits in Tables~\ref{tab_lines} and \ref{tab_continuum}. The dust continuum is clearly detected in all the frequency settings. We consider lines detected if the best fit of the line flux (as inferred from the median value of the posterior distribution) exceeds its 3-$\sigma$ confidence level. For non-detections, we use a line flux upper limit based on the 3-$\sigma$ confidence level of the posterior distribution, under the assumption of a fixed line width (375\,\kms{}, based on \Cii{}). All the fine structure lines in our study, as well as the CO(7--6) line, the OH doublet, and the H$_2$O 3(3,1)--3(2,2) and 3(0,3)--2(1,2) transitions match this criterion, whereas the J$_{\rm up}\geq 15$ transitions from CO, and the transitions involving the J$_{\rm up}\geq 4$ levels from H$_2$O remain undetected, and will be considered as 3-$\sigma$ upper limits in our analysis. We note that, for the sake of internal consistency, we refit the \Cii{} line with the same approach as all of the other lines analyzed in this paper, rather than referring to values reported in the literature.

We derive line luminosities as:
\begin{equation}\label{eq_linelum}
\frac{L_{\rm line}}{\rm L_\odot}=\frac{1.04\times10^{-3}}{1+z} \, \frac{F_{\rm line}}{\rm Jy\,km\,s^{-1}} \, \frac{\nu_0}{\rm GHz} \, \left(\frac{D_{\rm L}}{\rm Mpc}\right)^2
\end{equation}
where $F_{\rm line}$ is the line flux, as measured by integrating over the fitted Gaussian profile, $\nu_0$ is the rest-frame frequency of the transition, and $D_{\rm L}$ is the luminosity distance; and
\begin{equation}\label{eq_linelprime}
\frac{L'_{\rm line}}{\rm K\,km\,s^{-1}\,pc^2}=\frac{3.25\times10^{7}}{1+z} \, \frac{F_{\rm line}}{\rm Jy\,km\,s^{-1}} \, \left(\frac{\nu_0}{\rm GHz}\right)^{-2} \, \left(\frac{D_{\rm L}}{\rm Mpc}\right)^2
\end{equation}
\citep[see][for a discussion of line luminosity definitions]{carilli13}.

\begin{figure}
\begin{center}
\includegraphics[width=0.99\columnwidth]{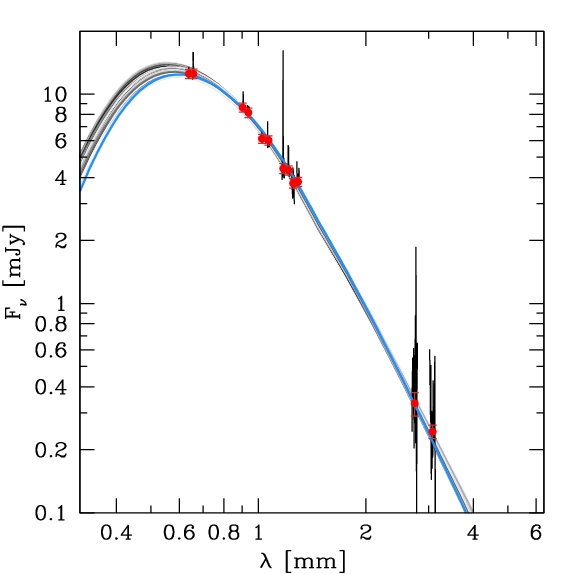}\\
\end{center}
\caption{Observed dust spectral energy distribution in PJ183+05. We plot the observed spectra and best-fit continuum flux density estimates from the spectral analysis as black lines and red points with error bars, respectively. The best fit model is shown in blue, while a random subset of the models within the 1-$\sigma$ confidence level are shown in gray for reference. The broad range of frequencies sampled in our study allows us to precisely pin down the dust temperature in this source.}
\label{fig_sed}
\end{figure}

\begin{table}
\caption{Continuum flux density measurements.}\label{tab_continuum}
\vspace{-5mm}
\begin{center}
\begin{tabular}{ccc}
\hline
Obs.Freq.  & $\lambda$  & $S_{\nu}$\\
GHz        & [mm]       & [mJy]    \\
(1)        & (2)        & (3)      \\
\hline
$ 97.435$  & $3.077$ & $0.244\pm0.028$ \\	    
$109.342$  & $2.742$ & $0.334\pm0.051$ \\	    
$231.946$  & $1.293$ & $3.81\pm0.38 $ \\	   
$239.520$  & $1.252$ & $3.75\pm0.38 $ \\	   
$246.646$  & $1.215$ & $4.33\pm0.44 $ \\	   
$255.459$  & $1.174$ & $4.42\pm0.45 $ \\	   
$280.875$  & $1.067$ & $5.82\pm0.58 $ \\	   
$292.989$  & $1.023$ & $6.12\pm0.61 $ \\	   
$319.788$  & $0.937$ & $8.15\pm0.82 $ \\	   
$331.688$  & $0.904$ & $8.63\pm0.86 $ \\	   
$457.035$  & $0.656$ & $12.52\pm1.26 $ \\	    
$469.093$  & $0.639$ & $12.51\pm1.26 $ \\	    
\hline
\end{tabular}
\end{center}
\end{table}

\subsection{Dust continuum modeling}\label{sec_continuum}

\begin{figure}
\begin{center}
\includegraphics[width=0.99\columnwidth]{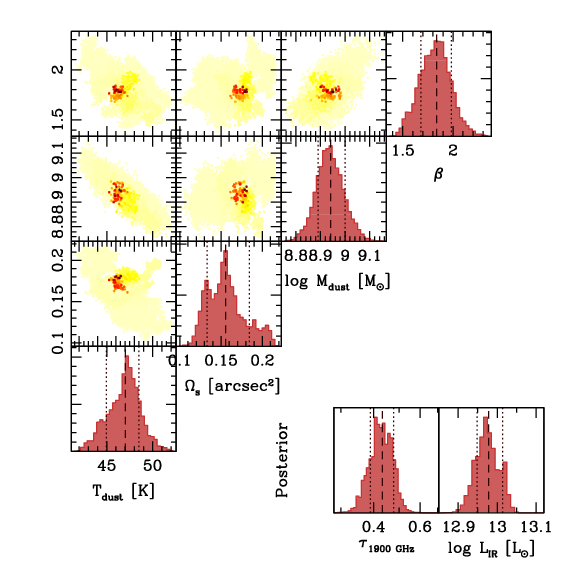}\\
\end{center}
\caption{Corner-plot posterior distributions for the parameters used in the dust SED modeling: the dust temperature, $T_{\rm dust}$; the observed solid angle of the emitting region, $\Omega_{\rm s}$; the mass of dust, $M_{\rm dust}$; and the dust emissivity index, $\beta$. In the corner plots, darker points show values corresponding to higher posterior probabilities. We also show the marginalized posterior distributions for the dust optical depth at the frequency of \Cii{}, $\tau_{\rm 1900\,GHz}$; and of the infrared luminosity, $L_{\rm IR}$. In all the marginalized distributions, the median values are marked with a dashed line, and the values corresponding to the 1-$\sigma$ confidence intervals are marked with dotted vertical lines.}
\label{fig_sed_pars}
\end{figure}

A modified black body provides us with a good description of the dust continuum in PJ183+05. Under the assumption of a single dust temperature, $T_{\rm dust}$, the flux density at a given observed frequency $\nu=\nu_0/(1+z)$ results from the black body emissivity, $B(\nu_0,T_{\rm dust})$ observed against the cosmic microwave background (CMB) at the quasar's redshift, integrated over the source apparent area, $\Omega_{\rm s}$, and corrected for the effects of radiative transfer:
\begin{equation}\label{eq_continuum}
S_\nu=\frac{\Omega_{\rm s}}{(1+z)^{3}}\,[B(\nu_0,T_{\rm dust}) - B(\nu_0,T_{\rm CMB})]\,[1 - \exp(-\tau_{\nu_0})]
\end{equation}
where:
\begin{equation}\label{eq_blackbody}
B(\nu_0,T)=\frac{2\,h\,\nu_0^3}{c^2}\,\frac{1}{\exp\left(\frac{h\nu_0}{k_{\rm b}T}\right)-1}
\end{equation}
is the black body emissivity law, $h$ is the Planck constant, $c$ is the speed of light, and $k_{\rm b}$ is the Boltzmann constant. The temperature of the CMB is $T_{\rm CMB}$ = 2.725\,(1+$z$)\,K = 20.27\,K at the quasar's redshift. The radiative transfer correction consists of a transmissive term and an absorbed term, where $\tau_{\nu_0}$ is the optical depth:
\begin{equation}\label{eq_tau}
\tau_{\nu_0}=\frac{\kappa(\nu_0)\,M_{\rm dust}}{\Omega_{\rm s} D_{\rm A}^2}.
\end{equation}
Here $M_{\rm dust}$ is the dust mass seen within a solid angle $\Omega_{\rm s}$; $D_{\rm A}=D_{\rm L}\,(1+z)^{-2}$ is the angular diameter distance; and $\kappa(\nu_0)$ is the dust emissivity law, for which we adopt an interpolation of the values reported in Table~5 of \citet{draine03} at $\nu_0>1500$\,GHz, and a power-law extrapolation $\kappa(\nu_0)=6.37\,(\nu_0/1500\,$GHz$)^\beta \,{\rm cm^2\,g^{-1}}$ at lower frequencies. In this framework, the overall dust continuum is uniquely determined by a combination of four parameters: $T_{\rm dust}$, $\Omega_{\rm s}$, $M_{\rm dust}$, and $\beta$, which we determine via a Monte Carlo Markov Chain fit. 
As priors, we adopt a log-normal distribution in $\tilde{T}_{\rm dust}=T_{\rm dust}-T_{{\rm CMB}}$ with scale parameter $\tilde{T}_{\rm ref}= $100\,K $-T_{{\rm CMB}}$ for the dust temperature (for reference, typical dust temperatures in quasar host galaxies at high redshifts are in the range $T_{\rm dust}= 40-100$\,K; see, e.g., \citealt{walter22}). For the angular size, $\Omega_{\rm s}$, we adopt a Gaussian distribution around the spatial extent reported in \citet{venemans20} (deconvolved dust sizes: $0.45''\times0.35'' \approx 2.4\times 1.9$\,kpc$^2$). The width of the prior distribution is set to 30$\%$ of the central value. We also adopt a log-normal distribution for the prior on the dust mass, centered around $M_{\rm dust}=10^8$\,\Msun{} \citep[based on the values typically found in $z>6$ quasars, see, e.g.,][]{venemans18} and with a 1-$\sigma$ width of 1\,dex; and a log-normal distribution centered on $\beta=1.5$ with a 1-$\sigma$ width of $0.1$\,dex \citep[see, e.g.,][]{beelen06}. We find $T_{\rm dust}=47.0_{-2.0}^{+1.5}$\,K, $\Omega_{\rm s}=0.155_{-0.022}^{+0.029}$\,arcsec$^2$ ($\approx 4.67_{-0.66}^{+0.87}$\,kpc$^2$), log $M_{\rm dust}$/\Msun{} = $8.94_{-0.05}^{+0.06}$, and $\beta=1.84_{-0.16}^{0.15}$.
Fig.~\ref{fig_sed} shows the best-fit model. Fig.~\ref{fig_sed_pars} shows the posterior distributions of the fitted parameters. The most identifiable correlation relates the dust mass and temperature, as expected in the optically-thin regime (from eq.~\ref{eq_continuum} and \ref{eq_tau}, for $\tau \ll 1$). 

We also infer posterior distributions for derived parameters, namely the optical depth at the \Cii{} frequency, $\tau_{\rm 1900\,GHz}$, and the IR luminosity, $L_{\rm IR}$. Eq.~\ref{eq_tau} gives us $\tau_{\rm 1900\,GHz}$. The IR luminosity is computed by integrating the dust emission model in the range IR luminosity in the 8--1000 $\mu$m wavelength range. We find an optical depth at the frequency of \Cii{} of $\tau_{\rm 1900\,GHz}=0.48\pm0.04$, and an IR luminosity of $\log L_{\rm IR}$\,[\Lsun]=$12.98_{-0.03}^{+0.04}$. The posterior distributions for $L_{\rm IR}$ and $\tau_{\rm 1900\,GHz}$ are also shown in Fig.~\ref{fig_sed_pars}.

We define the colors of the best-fit template of the dust emission as the ratios between the rest-frame monochromatic luminosities of the dust at different wavelengths. Namely, for PJ183+05 we measure $C(60/100)=0.84$ and $C(88/122)=1.4$ for the colors between $\lambda_0$=60 and 100 $\mu$m and 88 and 122 $\mu$m, respectively.

\subsection{Emission line modeling}\label{sec_emissionlines}

In order to put our observations in the context of the physical properties of the interstellar medium in PJ183+05, we derive analytical predictions based on first principles, and complement them with grids of predicted line ratios using the photoionization code \textsf{Cloudy} \citep{ferland98,ferland13,ferland17}.
We treat all the emission lines as arising from three distinct components (ionized, neutral, molecular), each described with a simple parametrization (e.g., constant density throughout each cloud). This is clearly a crude approximation that does not capture the complex structure of the ISM in galaxies. E.g., both spatial and velocity shifts have been reported among \Oiii{} and \Cii{} emission in high-$z$ galaxies \citep[e.g.,][]{carniani18}. However, these approximations suffice in order to infer luminosity--weighted properties of the galaxy as a whole.

\subsubsection{Analytical prescriptions from first principles}\label{sec_model}

Analytical prescriptions and empirical scaling relations can provide us with a back-of-the-envelope approach to the radiative transfer problem. They rely on a number of simplifications and assumptions (e.g., photoionization equilibrium, local thermal equilibrium) that are likely not universally valid when one studies the ISM of a galaxy in detail. However, this method allows for a straightforward interpretation of the observed line ratios in terms of physical quantities. Analytical implementations of the radiative transfer in ionized bubbles can work remarkably well, when compared with more sophisticated \textsf{Cloudy} models \citep[see, e.g.,][]{yang20_oiii,lamarche17,lamarche22}. 

In order to predict the luminosity of the ionized gas transitions in our study, we assume that a fully--ionized gas (i.e., volume density $n=n_e$) is in local thermodynamical and photoionization equilibrium, with $T_{\rm gas}$ setting the energy distribution of the electrons following a Maxwell distribution. Free electrons are the main collision partners in a fully--ionized gas cloud. Bound electrons are excited to energy levels beyond ground state via collisions, and are de-excited via collisions and radiation. The population of level\footnote{Following the literature \citep[e.g.,][]{draine11}, we use the indexes $i,j$ for pairs of energy levels in no particular energy order, and the indexes $l,u$ whenever it will be convenient to distinguish ``lower'' and ``upper'' energy levels.} $i$ is thus set by solving a system of linear equations of the form:
\begin{equation}\label{eq_radtransf}
\frac{d n_i}{dt}=\sum_{j\neq i}\,R_{ji}n_j - n_i\,\sum_{j\neq i} R_{ij} = 0.
\end{equation}
Here $R_{ul}$ are the rates at which electrons are de-excited from a higher-energy level $u$ to a lower-energy level $l$:
\begin{equation}\label{eq_Rul}
R_{ul} = n k_{ul}(T_{\rm gas})+(1+n_{\gamma, ul})\,A_{ul},
\end{equation}
where $n_\gamma$ is the photon occupation fraction, $n_\gamma = c^3\,(8 \pi \,h\nu^3)^{-1}\,u_\nu$, and $u_\nu$ is the energy volume density of the radiation field. On the other hand, $R_{lu}$ are the rates at which electrons are excited from $l$ to $u$:
\begin{equation}\label{eq_Rlu}
R_{lu}=\frac{g_u}{g_l}\,\left[n k_{ul}\,\exp\left(-\frac{E_{ul}}{k_{\rm b}T_{\rm gas}}\right)+n_{\gamma,lu}\,A_{ul}\right].
\end{equation}
Here $A_{ul}$ are the Einstein coefficients, $g_{i}$ are the statistical weights of the levels, $k_{ul}(T_{\rm gas})$ are the collision rates:
\begin{equation}\label{eq_kul}
k_{ul}(T_{\rm gas})\approx \frac{8.629\times 10^{-6}}{\rm [cm^{-3}\,s^{-1}\,K^{1/2}]} \, \frac{\Omega(l,u)}{g_u \,\sqrt{T_{\rm gas}}}. 
\end{equation}
and $\Omega(l,u)$ are the collision strengths, which we take from Appendix F of \citet{draine11}. The energy difference between two energy levels is $E_{ul}=h\nu_{ul}$. We assume that $n_\gamma$ is negligible. For 6-electron ions (C, N$^{+}$, O$^{++}$), we consider a five-level structure where electrons populate the $^3 P_0$ (ground), $^3P_1$, $^3P_2$, $^1D_2$, or $^1S_0$ levels, corresponding to $g_i$=(2\,J+1)=1, 3, 5, 5, and 1, respectively. For 5-electron ions (specifically, C$^+$), we only consider the two lower energy levels, $^2P_{1/2}$ (ground) and $^2P_{3/2}$, as the next levels ($^4P_{1/2}$, $^4P_{3/2}$, $^4P_{5/2}$) are only significantly collisionally populated at $T_{\rm gas}\gsim 50000$\,K.
\begin{figure}
\begin{center}
\includegraphics[width=0.99\columnwidth]{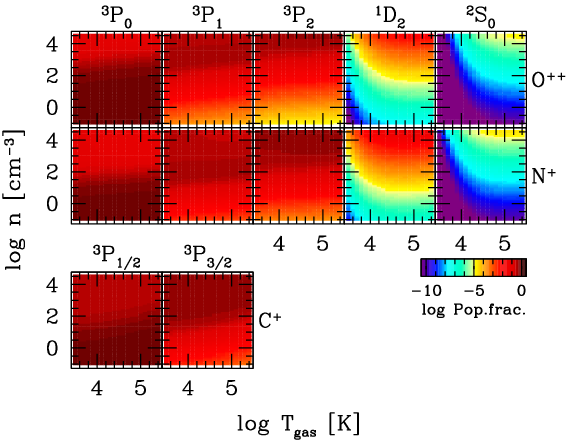}
\end{center}
\vspace{-5mm}
\caption{Results from the analytical model of the fine-structure line emission associated with O$^{++}$, N$^+$ and C$^+$ ions in the ionized medium, as described in Sec.~\ref{sec_cloudy}. Energy levels are populated by collisions with electrons, and depopulated by collisions and by radiative de-excitation. We assume local thermal equilibrium, thus the number of collisions depends only on the gas temperature $T_{\rm gas}$ and density $n$. The vast majority of electrons reside in the $^3P_{\rm J}$ levels. 
}
\label{fig_fsl_model}
\end{figure}

From our analysis, we infer the critical densities of the various levels, i.e., the densities at which collisional de-excitation equals radiative de-excitation:
\begin{equation}\label{eq_ncrit}
n_{{\rm crit}, i}=\frac{\sum_{j<i}\,A_{ij}}{\sum_{j\neq i} k_{ij}(T_{\rm gas})}.
\end{equation}
Tab.~\ref{tab_ncrit} lists the critical densities thus computed for the relevant transitions.

\begin{table}
\caption{Critical densities for the main far--IR fine-structure lines from the ionized gas considered in our work, computed for different gas temperatures $T_{\rm gas}$. For \Cii, we consider the ionized component only.}\label{tab_ncrit}
\vspace{-5mm}
\begin{center}
\begin{tabular}{cc|ccc}
\hline
Ion                & Line  & \multicolumn{3}{c}{$T_{\rm gas}$ [K]} \\
        	   &       & 5,000 & 10,000 & 20,000 \\
(1)     	   & (2)   & (3)   & (4)    & (5)    \\
\hline
\Nii{}  	   & 1-0   &  127  &  169   &  222   \\
\Nii{}  	   & 2-1   &  199  &  260   &  333   \\
\Oiii{} 	   & 1-0   & 1307  & 1798   & 2496   \\
\Oiii{} 	   & 2-1   & 2227  & 3048   & 4163   \\
\Cii{}$^{\rm ion}$ &3/2-1/2&   41  &   54   &	71   \\
\hline
\end{tabular}
\end{center}
\end{table}

Solving the system of equations \ref{eq_radtransf} as a function of $n$ and $T_{\rm gas}$ allows us to determine the fraction of the electron population in each level. This is shown in Fig.~\ref{fig_fsl_model}. The line luminosity associated with an optically--thin transition is then:
\begin{equation}\label{eq_Lc_model}
L_{ul}=n_u \, A_{ul} \,h\nu_{ul} V,
\end{equation}
where $V$ is the total volume in H{\sc ii} regions. Compared to the analysis of, e.g., \citet{yang20_oiii}, we assume that all the ionizing photons contribute to the photoionization equilibrium, and that the sizes of the bubbles where O$^{++}$, N$^{+}$ and H$^+$ reside are comparable. In sec.~\ref{sec_results} we will discuss the impact of these assumptions.

We assume that photoionization sets the abundance of ions. For a given template of the ionizing source, we can derive the number of photons emitted by the photoionizing source per unit time that are energetic enough to ionize different elements and ions. Namely, for a species X$^{N}$ with $N$ marking the ionization state (0=neutral, 1=first ionization, 2=second ionization, etc), we have that: 
\begin{equation}\label{eq_Qdef}
Q({\rm X}^{N})=\int_{\nu_{\rm ion}}^{\infty} \frac{L_\nu}{h \nu} d \nu
\end{equation}
and we define its net equivalent as the difference between the number of photons emitted per unit time that can photoionize X$^N$ into X$^{N+1}$ minus the photons that can photoionize X$^{N+1}$ into X$^{N+2}$:
\begin{equation}\label{eq_netQdef}
Q^{\rm net}({\rm X}^{N})=Q({\rm X}^{N})-Q({\rm X}^{N+1}).
\end{equation}

\begin{figure}
\begin{center}
\includegraphics[width=0.99\columnwidth]{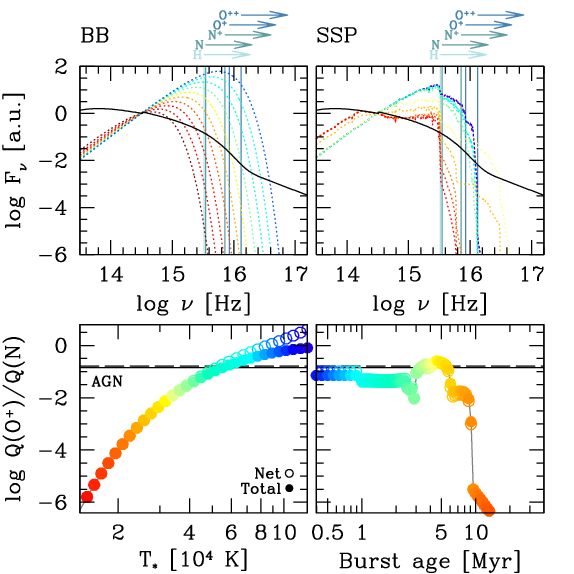}\\
\end{center}
\caption{Template comparison for the photoionizing source. {\em Top panels:} The rest-frame flux density of the AGN template (solid black lines), of black bodies with different temperatures $T_*$ (dotted lines in the left-hand panel) and single stellar populations with different ages (dotted lines in the right-hand panel). The frequency threshold of photons responsible for the ionization of hydrogen, nitrogen, oxygen relevant to this study are marked for reference. {\em Bottom panels:} The ratio of O$^+$ to N-ionizing photons produced by different templates, as a function of $T_*$ (black bodies; left-hand panel) and burst age (right-hand panel). Empty circles refer to values integrated at any $\nu_0>\nu_{\rm ion}$, while filled circles refer to the `net' values (see eq.~\ref{eq_netQdef}).
}
\label{fig_templates}
\end{figure}

Fig.~\ref{fig_templates} compares the templates used in our radiative transfer analysis, and their yield in terms of flux of photons energetic enough to singly and doubly ionize nitrogen, and doubly and triply ionize oxygen. As the source of photoionization we consider either a black body with varying temperature $T_*$ (mimicking the impact of individual massive stars), an AGN, or a single stellar population. For the AGN case, we adopt the default template in \textsf{Cloudy} \citep[based on][]{zamorani81,francis93,elvis94}:
\begin{equation}\label{eq_agn}
F_{\nu_0}=\nu_0^{-0.5}\,\exp\left(-\frac{h \nu_0}{k_{\rm b} T_{\rm BB}}\right) \,\exp\left(-\frac{k_{\rm b} T_{\rm IR}}{h \nu_0}\right) + a\, \nu_0^{-1}
\end{equation} 
with the IR cutoff determined by $k_{\rm b} T_{\rm IR}=0.136$ eV or $T_{\rm IR}$=1580 K; the Blue Bump temperature set to $T_{\rm BB}=1.5\times10^5$ K; and $a=2.4\times10^6$ is a coefficient set to define the relative strength of the Blue Bump and the X-ray corona, which in our case is constrained by the requirement that $\alpha_{\rm ox}=-1.4$. For the single stellar population model, we refer to \citet{bruzual03} models with solar metallicity and for a Salpeter initial stellar mass function, computed at various time steps after the initial burst. The hardness of the radiation field from these templates, gauged by $Q({\rm O^+})/Q({\rm N})$, increases with $T_*$ for the black body templates, reaching the AGN value for $T_*\approx 50000$\,K. The single stellar population shows a non-monotonic behavior, due to the rapid evolution of the most massive stars from the main sequence to the asymptotic giant branch phase; but we find an overall trend, with the $Q({\rm O^+})/Q({\rm N})$ ratio slowly decreasing in the first $\sim 10$\,Myr since the burst, followed by a rapid decline as all massive (O, B-type) stars evolve out of the main sequence.

The analytical model used for the ionized medium cannot be trivially expanded to the PDR/XDR regime. In this regime, collision excitation/de-excitation terms need to account for multiple collision partners (electrons, neutral H atoms, para-H$_2$, ortho-H$_2$, He, etc), the abundance of which depends on the ionization and dissociation conditions in different layers of the clouds. Furthermore, the chemistry itself of the neutral and molecular phases is more complicated (e.g., photo-dissociation of carbon monoxide molecules alters the abundance of neutral carbon atoms in the clouds). Opacity is often non-negligible (both for lines and continuum). Finally, many parameters are interconnected (e.g., the kinetic temperature of the gas, $T_{\rm gas}$, might be partially coupled with the dust temperature, $T_{\rm d}$). For instance, \citet{harrington21} build on the formalism outlined in \citet{weiss07} to solve the radiative transfer problem simultaneously for the dust and the molecular gas. Their modeling involves seven free parameters (gas density, $n$; gas kinetic temperature, $T_{\rm gas}$; dust temperature, $T_{\rm dust}$; turbulence velocity; virial velocity gradient; carbon abundance, [C/H$_2$]; scale size of the emitting region) for each gas component. Other parameters, such as the [CO/H$_2$] abundance of the gas-to-dust mass ratio, are set independently. Their approach allows them to characterize various physical properties of the dust and molecular gas in a sample of Planck--selected sub-mm galaxies. While the method by \citet{harrington21} can in principle be applied here, our observational constraints (in particular concerning the CO spectral line energy distribution) are too sparse to lead to informative results. For the analysis of the neutral and molecular regimes, we thus opt to stick to the \textsf{Cloudy} models presented in \citet{pensabene21} and summarized in the next section.

\subsubsection{Cloudy modeling}\label{sec_cloudy}

We based our \textsf{Cloudy} analysis on the models presented in \citet{pensabene21} and \citet{meyer22}. Namely, we model the ISM as a homogeneous plane-parallel slab of gas exposed to a radiation field. We independently model the ionized vs.\ neutral and molecular phases.

For all the input templates of photoionization sources, we fix the the ionization parameter $U$, which represents the density of ionizing photons per hydrogen atom:
\begin{equation}\label{eq_U}
U=\frac{Q({\rm H})}{4 \pi\,r^2\,n\,c},
\end{equation}
where $Q({\rm H})$ is the number of hydrogen-ionizing photons emitted per unit time by the photoionizing source, and $r$ is the distance between the photoionizing source and the cloud. \citet{rigopoulou18} use the dust color $C(88/122)$ as a proxy for the ionization parameter. Following their approach, the measured $C(88/122)\approx1.4$ corresponds to log\,$U=[-1.9,-2.5]$ for a gas density $n$=$[10,1000]$\,cm$^{-3}$, respectively (see their Fig.~4). As we aim to compare the emission of lines with different ionization energies close to or above the ionization energy of hydrogen, for this part of the analysis it is more critical to study the hardness of the radiation field at energies higher than the ionization energy of hydrogen. We thus use a fixed log\,$U$=$-2$ in our analysis. 

We assume that the gas volume density is constant within the cloud, and we sample the line emissivity over a range of $\log n$\,[cm$^{-3}$] $=[0, +3]$ in steps of $0.25$\,dex. We assume ISM and PAH grains. The metallicity\footnote{Throughout the paper, we implicitly refer to the metallicity and abundances of the gas phase only, defined as Z=$\log{\rm [O/H]}+12$. We refer the interested reader to \citet{nicholls17} for prescriptions on how to map these quantities to the equivalent stellar ones.} in our grids ranges between $\log Z$\,[Z$_\odot$] $=[-1, +0.4]$ in steps of 0.2\,dex. Our measurements are sensitive to the relative abundance of elements rather than to metallicity per se. Various works have addressed how to infer metal abundances and metallicities based on far-IR FSL transitions, occasionally aided by observations of nebular rest-frame optical lines \citep[e.g.,][]{peng21,lamarche22}. Here, following \citet{nicholls17}, for the abundance of carbon, nitrogen, and helium we adopt:
\begin{equation}\label{eq_C_O}
\log {\rm [C/O]}=\log \left(10^{-0.8}+10^{\log {\rm [O/H]}+ 2.72}\right)
\end{equation}
\begin{equation}\label{eq_N_O}
\log {\rm [N/O]}=\log \left(10^{-1.732}+10^{\log {\rm [O/H]}+ 2.19}\right)
\end{equation}
\begin{equation}\label{eq_He_H}
\log {\rm [He/H]}=-1.0783 + \log \left(1+0.1703\,\frac{\rm Z}{\rm [Z_\odot]}\right),
\end{equation}
while we use the default ISM abundances in \textsf{Cloudy} and a linear scaling with [O/H] for all the other elements.

We also include a CMB background computed at the source's redshift, and a turbulence with an RMS of 100 \kms{}. We stop the integration when we reach a ionized fraction of $x_{\rm HI}=0.05$.

For the ionized medium, we run \textsf{Cloudy} for all the templates shown in Fig.~\ref{fig_templates}. For the neutral and molecular phases, we adopt the models of Photon-Dominated Regions (PDRs) and X-ray Dominated Regions (XDRs) presented in \citet{pensabene21}. In brief, we compute the radiative transfer using \textsf{Cloudy} for a grid of $15\times 18\times 3$ combinations of gas volume density, intensity of the radiation field, and column density. For the former, we adopt $\log n$\,[cm$^{-3}$] = [2, 6] (0.29 dex spacing). For the latter, we adopt a range of $\log G_0$ = [1, 6] (0.29 dex spacing) for PDRs, and $\log F_{\rm X}$ [erg\,s$^{-1}$\,cm$^{-2}$] = [-2.0, 2.0] (0.24 dex spacing) for XDRs. Contrary to what we have done for the ionized medium, we here fix the shape of the stellar template using a black body with $T_*$=50,000\,K, as the intensity of the radiation field is more critical in setting the PDR conditions. For the AGN, we adopt the same template as in eq.~\ref{eq_agn}. We adopted the default \textsf{Cloudy} recipes for the cosmic ray ionization rate background. 

The total cloud column density lies in the range $\log N_{\rm H}$\,[cm$^{-2}$]=[22, 24] (1 dex spacing). A higher column density $N_{\rm H}>5\times 10^{23}$ cm$^{-2}$ is required to properly model H$_2$O and OH emission \citep[e.g.,][]{goicoechea05,goicoechea06,gonzalezalfonso14,pensabene22}. However, because of the sparsity of the water and hydroxyl transitions studied here, an extensive analysis of the radiative properties of these molecules is beyond the scope of this work.

\section{Results}\label{sec_results}

In this section, we use the observed line luminosities to infer physical properties of the ISM.  

\subsection{Opacity}\label{sec_opacity}

If the line-emitting clouds are interspersed within the dusty medium, line emission should be corrected for opacity. In sec.~\ref{sec_continuum}, we inferred a non-negligible opacity, $\tau_\nu\approx 0.47$ at the frequency of \Cii{}. For this opacity, the flux density emerging at $\nu_0\sim 1900$\,GHz is 2/3 of the intrinsic value. The higher the frequency, the larger the correction (see eq.~\ref{eq_tau}), with the largest value amounting to 4.6 for \Oiii{}. The extinction values computed at the frequencies of the lines studied in this work are listed in Tab.~\ref{tab_lines}. 

If we correct for opacity by scaling all the line luminosities by a factor $e^{\tau_\nu}$, we find that the intrinsic \Nii{}$_{\rm 122\,\mu m}$/\Nii{}$_{\rm 205\,\mu m}$ ratio would be 1.77$\times$ higher than the observed value, while the intrinsic \Oiii{}$_{\rm 88\,\mu m}$/\Nii{}$_{\rm 122\,\mu m}$ would be 2.05$\times$ higher than the observed value. Our measurements for the molecular phase would be virtually unaltered, given that they mostly rely on lower-frequency lines. As we will show in the following subsections, applying these corrections would result in a higher estimate of the electron density ($n>200$\,cm$^{-3}$) and a slightly harder radiation field. However, the actual values of such corrections strongly depend on how the line--emitting gas clouds and the dust are distributed along the line of sight, which is unknown. Because of this uncertainty, and due to the modest impact that this correction would have on our final results, we opt not to apply the opacity correction to our fiducial measurements.

\subsection{Inferred \Nii{}$_{205 \mu{\rm m}}$ and the origin of \Cii{}}\label{sec_fpdr}

We estimate the luminosity of the \Nii{}$_{\rm 205\,\mu m}$ line following the empirical trend presented in \citet{lu15} for a sample of local luminous IR galaxies:
\begin{equation}\label{eq_lu}
\log {\rm [NII]/[CII]} = (-0.65\pm0.08)\,C(60/100)-(0.66\pm0.06)
\end{equation}
where the \Nii{}$_{205\,\mu{\rm m}}$ to \Cii{}$_{158\,\mu{\rm m}}$ line ratio is tied to the color of the dust emission, $C(60/100)$, computed as the ratio between the rest-frame flux density at 60 and 100 $\mu$m. In the case of PJ183+05, $C(60/100)=0.84$, yielding $\log L_{\rm [NII]}$\,[\Lsun]$\approx 8.51\pm0.23$, where the uncertainty is dominated by the scatter in the relation from \citet{lu15}. The \Cii{}$_{158\,\mu{\rm m}}$ and \Nii{}$_{205\,\mu{\rm m}}$ lines both have relatively low critical density (see Tab.~\ref{tab_ncrit}, assuming electrons as collision partners) and have similar ionization energies (14.5\,eV for nitrogen, 11.2\,eV for carbon). In the ionized medium, their luminosity ratio is thus determined almost exclusively by the abundance ratio, [C$^+$/N$^+$]. This is confirmed by both our \textsf{Cloudy} modeling and our analytical prescription, as shown in Fig.~\ref{fig_cii_nii_predict}. We find that, for the ionized component alone, the expected \Cii{}$_{158\,\mu{\rm m}}^{\rm ion}$/\Nii{}$_{205\,\mu{\rm m}}$ luminosity ratio is 6--12 at low densities, and 10--20 at high densities, with the spread completely dominated by the relative abundance. \citet{romano23} has recently reviewed the evolution of carbon, nitrogen and oxygen in galaxies. Here we refer to the \citet{nicholls17} calibration, which yields [C$^+$/H]=$2.6\times10^{-4}$ and [N$^+$/H]=$6.2\times10^{-5}$ at solar metallicities, or [C$^+$/N$^+$]=4.28. On the other hand, in the literature it is common to follow \citet{oberst06}, who rely on the abundance estimates in \citet{savage96}, [C$^+$/H]=$1.4\times10^{-4}$ and [N$^+$/H]=$7.9\times10^{-5}$. These slightly lower (higher) carbon (nitrogen) abundances result in a $\sim 2.4$ lower relative abundance, [C$^+$/N$^+$]=1.77. 

Because \Nii{} only arises from the ionized medium, while \Cii{} can trace both the ionized and the neutral/molecular medium in PDR/XDR regimes, we can use the expected \Cii{}$_{158\,\mu{\rm m}}^{\rm ion}$/\Nii{}$_{205\,\mu{\rm m}}$ luminosity ratio based on our modeling (see Fig.~\ref{fig_cii_nii_predict}) to estimate the fraction of the \Cii{} emission which arises from the PDR/XDR environment:
\begin{equation}\label{eq_cii_frac}
f({\rm [CII]^{PDR}}) = \frac{\rm [CII]^{PDR}}{\rm [CII]}\approx 1-9\, \frac{\rm [NII]}{\rm [CII]} \approx 0.44
\end{equation}
for a gas density of $n\sim 180$\,cm$^{-3}$ and solar metallicities. We point out that this value is highly sensitive to the assumed relative abundances. If we adopt the traditional values from \citet{savage96}, we find a much higher $f({\rm [CII]^{PDR}})=0.81$.

\subsection{Electron density and size of the H{\sc ii} regions}\label{sec_e_dens}

\begin{figure}
\begin{center}
\includegraphics[width=0.49\textwidth]{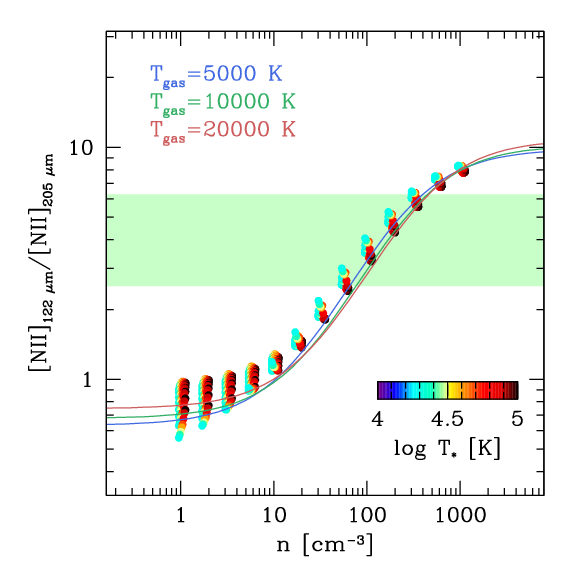}\\
\end{center}
\caption{Predicted \Nii{}$_{122\,\mu{\rm m}}$/\Nii{}$_{205\,\mu{\rm m}}$ luminosity ratio, as a function of the gas density. The results from our \textsf{Cloudy} models are shown as colored symbols, with small offsets introduced for the sake of clarity. These models are color--coded by the hardness of the impinging radiation field, parameterized as the effective temperature of black bodies $T_*$. We also show the predicted ratios for our analytical prescriptions as solid lines, color-coded by gas temperature $T_{\rm gas}$. The observed ratio is marked with green shading. The predicted ratio is practically independent of $T_*$ or $T_{\rm gas}$, and is solely determined by the gas density $n$.
}
\label{fig_fsl_density}
\end{figure}

The ratio between the two far-IR fine-structure lines of \Nii{} at 122\,$\mu$m and 205\,$\mu$m is independent of metallicity (as they are two transitions associated with the same ion). It is also insensitive to the hardness of the radiation field and to the gas temperature, due to the low energy of the $^3 P_1$ and $^3 P_2$ levels ($E/k_{\rm b}=70$ and 188 K, respectively) compared to the typical electron temperatures in the ionized medium ($T_{\rm gas}=5000$--20000\,K). On the other hand, the two transitions have different critical densities (see Tab.~\ref{tab_ncrit}). Thus, their luminosity ratio is sensitive to the electron density in the range 3--3000\,cm$^{-3}$. This is shown in Fig.~\ref{fig_fsl_density}, where again we show both the results from our \textsf{Cloudy} models and from our analytical prescriptions. The two approaches lead to consistent results throughout the range of interest. Using the measured \Nii{}$_{122\,\mu{\rm m}}$ luminosity from Tab.~\ref{tab_lines}, and the inferred \Nii{}$_{205\,\mu{\rm m}}$ from the previous section, we find a ratio of $\log$\Nii{}$_{122\,\mu{\rm m}}$/\Nii{}$_{205\,\mu{\rm m}}$=$0.56_{-0.16}^{+0.23}$, corresponding to a gas density of $\log n$ [cm$^{-3}$] $=2.26\pm 0.35$. 

Our density estimate allows us to compute a fiducial volume occupied by the ionized gas, by comparing the observed line luminosity with the population fraction and gas density estimated via models. For $T_{\rm gas}=10,000$\,K, $n$=180\,cm$^{-3}$, and Z=Z$_\odot$, using eq.~\ref{eq_Lc_model} in the case of the \Nii{}$_{122\,\mu{\rm m}}$ transition, we obtain an effective volume of the ionized gas, $V_{\rm HII}^{\rm eff}\sim 0.2$\,kpc$^3$. Assuming this volume is organized in a collection of $N^{\rm HII}$ spherical H{\sc ii} regions, the Stromgren radius would be $\sim 360\,(N^{\rm HII})^{-1/3}$\,pc. 

\subsection{Hardness of the radiation field and metallicity}

The first ionization energies of carbon, nitrogen and oxygen are 11.26, 14.53 and 13.61 eV, respectively. Their second ionization happens at 24.38 eV, 29.60 eV, and 35.11 eV, respectively. Finally, the third ionization energy of oxygen is 54.93 eV. Thus, the ratios \Oiii{}/\Cii{}$^{\rm ion}$ and \Oiii{}/\Nii{} are sensitive to the hardness of the photoionizing radiation, i.e., the relative number of photons produced per unit time in the band 35--55 eV vs.\ the band 14--25 eV (see Fig.~\ref{fig_templates}). In addition, these line ratios also depend on the relative abundance of the elements, and on the gas density (see Tab.~\ref{tab_ncrit}). Assuming the best-fit value for $n$ from sec.~\ref{sec_e_dens}, we can use the \textsf{Cloudy} models and analytical prescriptions described in sec.~\ref{sec_cloudy} to predict the \Oiii{}$_{88\,\mu{\rm m}}$/\Nii{}$_{122\,\mu{\rm m}}$ ratio as a function of the hardness of the radiation field, parameterized as the effective temperature of a black body, $T_*$, and of metallicity $Z$. We opt to focus on the \Oiii{}/\Nii{} ratio rather than on the \Oiii{}/\Cii{}$^{\rm ion}$ ratio because of the uncertainties on $f($\Cii$^{\rm PDR})$ discussed in sec.~\ref{sec_fpdr}.

\begin{figure}
\begin{center}
\includegraphics[width=0.49\textwidth]{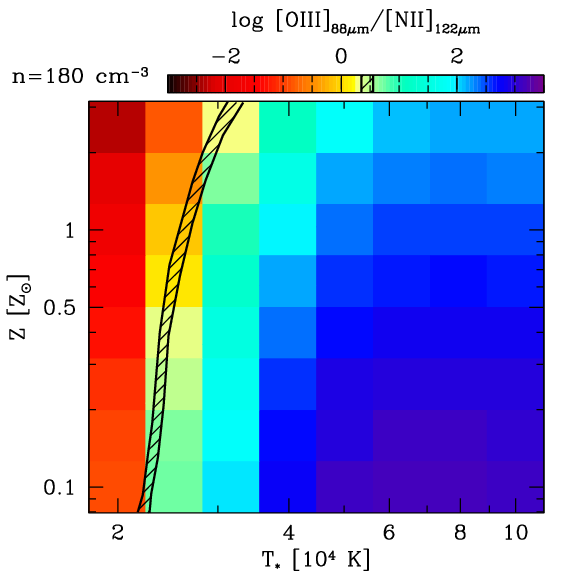}\\
\end{center}
\caption{The \Oiii{}$_{\rm 88\,\mu m}$/\Nii{}$_{\rm 122\,\mu m}$ luminosity ratio as a function of metallicity Z and hardness of the photoionizing radiation, parameterized as the temperature of an equivalent black body, $T_*$, computed from \textsf{Cloudy} models assuming a gas density of $n=180$\,cm$^{-3}$. The shaded area marks the observed line ratio and its 1-$\sigma$ confidence interval. The line ratio is nearly independent of metallicity at Z$\lsim$0.5\,Z$_\odot$ (see eq.~\ref{eq_N_O}), while it is strongly sensitive to the hardness of the ionizing radiation. Our observations point to a value of $T_*$=21,000--32,000\,K.
}
\label{fig_fsl_ratios}
\end{figure}

\begin{figure}
\begin{center}
\includegraphics[width=0.49\textwidth]{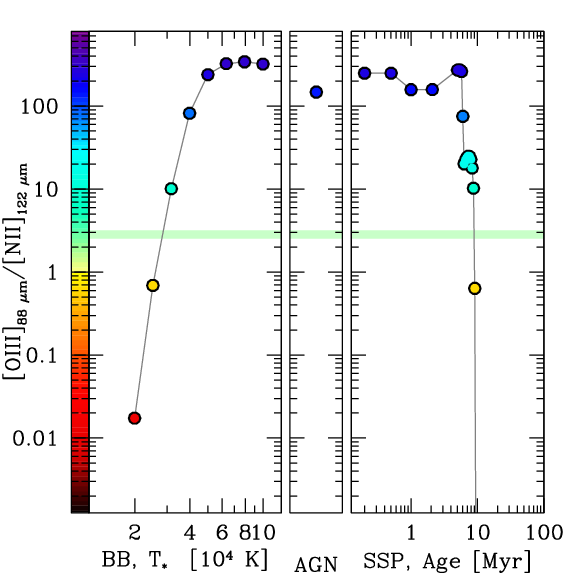}\\
\end{center}
\caption{The \Oiii{}$_{\rm 88\,\mu m}$/\Nii{}$_{\rm 122\,\mu m}$ luminosity ratio, computed with \textsf{Cloudy}, assuming $n=180$\,cm$^{-3}$ and Z=Z$_\odot$, for the various templates adopted in our study: A black body radiation of various temperatures, $T_*$; an AGN as described in eq.~\ref{eq_agn}; and a single stellar population with different burst ages, based on \citet{bruzual03}. The observed line ratio is marked with a light green shading. The observed line ratio is consistent with the expectations for a black body of $T_*\approx 25,000$\,K or with a single stellar population with age $\approx 9$\,Myr. The AGN template, as well as templates involving either hotter black body temperatures, or lower stellar population ages, appear to over-predict the \Oiii{}/\Nii{} luminosity ratio.
}
\label{fig_templ_comp}
\end{figure}

Our predictions are shown in Figs.~\ref{fig_fsl_ratios} and \ref{fig_templ_comp}. We find that the \Oiii{}$_{88\,\mu{\rm m}}$/\Nii{}$_{122\,\mu{\rm m}}$ ratio shows only a mild dependence on the gas metallicity. The evolution of the [N/O] relative abundance set by eq.~\ref{eq_N_O} suggests that the ratio is constant at low metallicities (Z$<$0.1\,Z$_\odot$, in the regime of primary abundances, where the enrichment is dominated by core-collapse supernovae) and becomes nearly linear with Z only at Z$>$Z$_\odot$ (where the secondary abundances arise due to delayed nucleosynthesis in intermediate-mass stars; see \citealt{nicholls17}). On the other hand, the \Oiii{}$_{88\,\mu{\rm m}}$/\Nii{}$_{122\,\mu{\rm m}}$ luminosity ratio is strongly sensitive to the hardness of the photoionizing radiation. In the case the input source is a black body, we find that that the expected line ratio changes by 2--3 dex for $T_*$ ranging between 20,000 and 40,000\,K. The AGN template yields a very high \Oiii{}$_{88\,\mu{\rm m}}$/\Nii{}$_{122\,\mu{\rm m}}$ luminosity ratio, $\sim 50$ times higher than observed. Finally, the single stellar population templates predict very high \Oiii{}$_{88\,\mu{\rm m}}$/\Nii{}$_{122\,\mu{\rm m}}$ luminosity ratios in the first $\sim 5$\,Myr, followed by a rapid decrease, with virtually no significant \Oiii{}$_{88\,\mu{\rm m}}$ expected $\sim 10$\,Myr after the burst. The observed line ratio points to a $T_*\approx 25,000$\,K for Z=Z$_\odot$, or a burst age of $\approx 9$\,Myr.

We notice that our \textsf{Cloudy} models and analytical prescriptions agree on qualitative but not quantitative terms (see appendix \ref{sec_model_vs_cloudy}). Because the two approaches are consistent in the other diagnostics discussed so far (involving $n$, $T_{\rm gas}$, and Z), we argue that the discrepancies arise due to the simplistic assumptions made in the analytical prescriptions concerning the budget of ionizing photons $Q$(N) and $Q$(O$^+$). 
That is, our prescription assumes that all of the photons contribute to the photoionization of nitrogen and oxygen. A more realistic description would consider the competing role of hydrogen photoionization within different layers of the cloud, and the frequency dependence of the cross-section of the process ($\propto \nu^{-3}$).
A more refined analytical description of the internal ionization structure of H{\sc ii} regions is possible \citep[see, e.g.,][]{yang20_oiii}. However, due to the steep dependence of the observed line ratio on $T_*$, our simple model for the H{\sc ii} regions results in a $T_*$ constraint that is only $\sim 1.7$ times higher than the one derived from the much more refined \textsf{Cloudy} simulations -- a discrepancy that has very little impact on the conclusions of our work. Thus, we argue that a more sophisticated model is unnecessary for the present work.

The observed \Oiii{}/\Nii{} luminosity ratio points to B-type stars as the main driver of the photoionization budget. These stars have a stellar mass of $\sim 15$\,\Msun{}, a luminosity of $\sim 10,000$\,\Lsun, and a main-sequence lifetime of $\sim 10$\,Myr. A quasar contribution to the photoionization budget, while likely present, is not required to explain the observed gas properties.

\subsection{PDR or XDR?}\label{sec_co_ci_cii}

\begin{figure}
\begin{center}
\includegraphics[width=0.99\columnwidth]{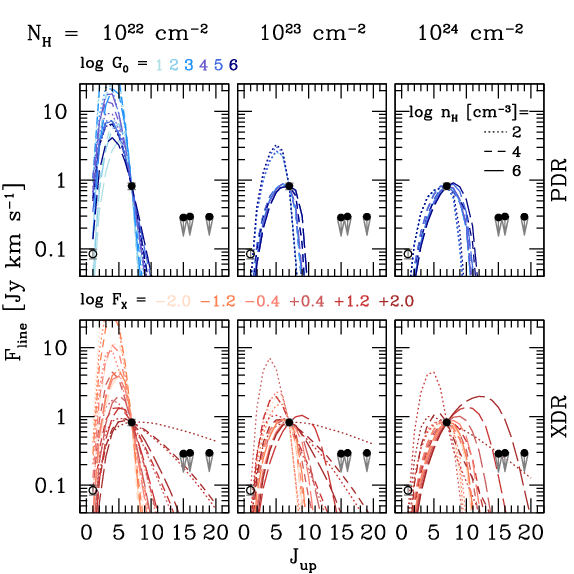}
\end{center}
\vspace{-5mm}
\caption{Observed constraints on the CO spectral line energy distribution in PJ183+05. Our measurements are shown as black filled circles. Top and bottom panels show the models for PDRs and XDRs, respectively. We mark models, taken from \citet{pensabene21}, with increasingly darker colors at increasing intensity of the radiation field. Dotted, short-dashed and long-dashed lines refer to models with gas density $\log n$ [cm$^{-3}$] = 2, 4, 6 respectively. All models are normalized to the observed CO(7-6) line flux. We plot the observed line fluxes or upper limits as black circles. In addition, we show the predicted CO(1-0) flux, as derived from the dust continuum, as described in sec.~\ref{sec_co_ci_cii}. Our constraints exclude a strong XDR environment, but leave room for a combination of PDR and XDR environments in the excitation conditions of the molecular ISM in PJ183+05.
}
\label{fig_co_sled}
\end{figure}

Our program encompasses four CO transitions, with J$_{\rm up}$=7, 15, 16, 19. Of these, only the CO(7-6) line is detected. This places loose constraints on the CO spectral line energy distribution (SLED; see Fig.~\ref{fig_co_sled}). We compare our observational results with the predictions for PDR and XDR models, as described in Sec.~\ref{sec_cloudy}. We also compute the expected line luminosity for the CO(1-0) transition, based on the dust mass estimate presented in sec.~\ref{sec_continuum}: $L'_{\rm CO(1-0)}=\alpha_{\rm CO}^{-1}\,\delta_{g/d}\,M_{\rm dust}$=$1.1\times10^{11}$\,\Msun{}. For the conversion, we assumed a gas-to-dust ratio of $\delta_{\rm g/d}$=100 \citep{berta16} and an $\alpha_{\rm CO}$=0.8\,\Msun{}\,(\Kkmspc)$^{-1}$ \citep{bolatto13}. Compared to the observed CO(7-6) luminosity, $L'_{\rm CO(7-6)}=(2.16\pm0.24)\times 10^{11}$\,\Kkmspc{},  this implies a line ratio $r_{71}=L'_{\rm CO(7-6)}/L'_{\rm CO(1-0)}$ = 0.20. This is in line with, e.g., the CO SLED template for $z>2$ galaxies in the ASPECS survey \citep[$r_{71}=0.17$;][]{boogaard20}.

The non-detections of the very high-J CO transitions in our sample only allow us to exclude the most extreme scenarios of pure XDR with strong incident radiation field, $F_{\rm X}\gsim 100$\,erg\,s$^{-1}$\,cm$^{-2}$ and $n>10^6$\,cm$^{-3}$. 

The combination of \Ci{}$_{\rm 370 \mu m}$ and \Cii{}$_{\rm 158 \mu m}$ provides a powerful diagnostic of the neutral and molecular medium. Models of XDRs predict that the harder X-ray radiation should be able to penetrate deeper into the cloud, and enhance the heating and the photo-dissociation of carbon monoxide compared to the predictions for models where the source of radiation is young stars. This should manifest as a low \Cii{}/\Ci{} ratio: $<10$ for a column density $N_{\rm H}$ = $10^{23}$ cm$^{-2}$, over a wide range of cloud density \citep[see, e.g.,][]{pensabene21}. \citet{decarli22} examined the ratio in a sample of $z\sim 6$ quasars, and found \Cii{}/\Ci{} ratios in the range 20--100, strongly pointing toward PDR rather than XDR regimes being the norm in the host galaxies of high--$z$ quasars. Here we measure a ratio of $\log$ \Cii{}/\Ci{} = $1.55_{-0.13}^{+0.17}$. Even if we consider only the component of \Cii{} that is not associated with the ionized medium, \Cii$^{\rm PDR}$, the ratio drops to $\sim 15$, still significantly higher than the predictions for XDRs. 

In summary, all these lines of evidence disfavor a scenario in which the neutral and molecular medium in PJ183+05 is dominated by XDR conditions, but leave room for a combination of PDR and XDR, or a PDR--dominated regime. A more extensive census of both lower- and higher-J CO transitions is required to further address the relative importance of PDR and XDR conditions in the host of PJ183+05.

\subsection{Water and hydroxyl emission}

\begin{figure*}
\begin{center}
\includegraphics[width=0.99\columnwidth]{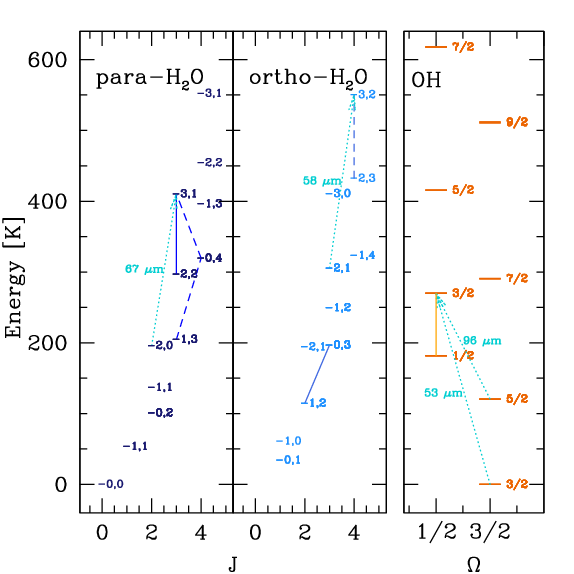}
\includegraphics[width=0.99\columnwidth]{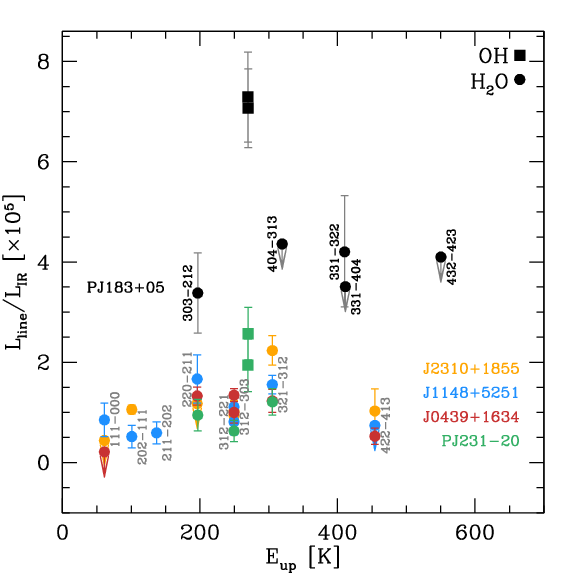}\\
\end{center}
\caption{{\em Left:} Scheme of the J$\leq$4 energy levels of the H$_2$O molecule, and for the $\Omega$=1/2 and 3/2 levels of the OH molecule. Solid lines mark the transitions detected in our study, dashed lines show the transitions that we observed but not detected. The channels of IR pumping relevant for this study are also marked with dotted arrows. {\em Right:} Comparison between the luminosity of H$_2$O and OH transitions, normalized to the dust IR luminosity, in PJ183+05 and in other $z>6$ quasars from \citet{pensabene21,pensabene22}. All of the water and hydroxyl transitions detected in our work show line-to-IR luminosities that are $\sim 3\times$ higher than the values observed in other quasars at the same redshift.}
\label{fig_water}
\end{figure*}

Water vapor (H$_2$O) has been detected in high redshift galaxies up to $z\sim 6.5$ \citep{vanderwerf11, riechers13, omont13, yang13, yang16, jarugula19, apostolovski19, yang19, li20, yang20, stacey20, stanley21, pensabene21, pensabene22, dye22, riechers22}. In our study, we detected two water transitions, 3(3,1)--3(2,2) from para-H$_2$O and 3(0,3)--2(1,2) from ortho-H$_2$O, while we do not detect the 4(3,2)--4(2,3), 4(0,4)--3(1,3) or 3(3,1)--4(0,4) transitions (see Fig.~\ref{fig_water}). To the best of our knowledge, this is the first time that the 3(3,1)--3(2,2) and 3(0,3)--2(1,2) lines have been detected beyond the local Universe. The para-H$_2$O 3(3,1) level is populated by collisions or via IR pumping at 67 $\mu$m. The energy level is then de-excited into 4(0,4) or 3(2,2) via collisions or radiative de-excitation. The ortho-H$_2$O 4(3,2) level is populated by the 58 $\mu$m pumping or via collisions. The non-detection of this transition sets a loose constraint on any mechanism populating the high-energy levels (J$>$3) of the water molecule (see Fig.~\ref{fig_water}).

Hydroxyl (OH) has also been detected in various low \citep[e.g.,][]{gonzalezalfonso14,gonzalezalfonso17,herreracamus20,runco20} and high redshift \citep[e.g.,][]{riechers13,riechers14,spilker18,spilker20,pensabene21} galaxies. In this work, we observe the OH $^2\Pi_{1/2}$ J=3/2$\rightarrow$1/2 transition at 163 $\mu$m. The upper level can be populated via IR pumping from the ground level of the $^2\Pi_{3/2}$ ladder (53\,$\mu$m), or via radiative de-excitation from the $^2\Pi_{1/2}$ J=5/2 level, which is populated via IR pumping at 35 $\mu$m (from ground level, $^2\Pi_{3/2}$ J=3/2), and at 48 $\mu$m \citep[from $^2\Pi_{3/2}$ J=5/2; see][and Fig.~\ref{fig_water}]{goicoechea06}. The non--zero electronic angular momentum of OH in its ground level introduces ``$\Lambda$''--doubling splitting of each energy level. In addition, ``hyperfine'' splitting occurs due to nuclear and electron spin coupling. The 163\,$\mu$m transition thus actually splits into a total of 6 transitions, at 
1834.7350\,GHz (N=$2^+$--$1^-$, F=1--1),
1834.7469\,GHz (N=$2^+$--$1^-$, F=2--1),
1834.7499\,GHz (N=$2^+$--$1^-$, F=1--0),
1837.7461\,GHz (N=$2^-$--$1^+$, F=1--1),
1837.8163\,GHz (N=$2^-$--$1^+$, F=2--1), and
1837.8365\,GHz (N=$2^-$--$1^+$, F=1--0). The hyper-fine structure splitting is negligible for typical extragalactic observations ($\lsim 0.1$\,GHz at rest frame, or $\lsim 15$\,\kms{}, i.e., much smaller than the width of the lines). Thus, the $\Lambda$--doubling implies that the 163\,$\mu$m transition appears as a doublet, spaced by $\approx 500$\,\kms{}. 

Fig.~\ref{fig_water}, {\em right} compares the line luminosity of the H$_2$O and OH transitions in PJ183+05, normalized to the total IR luminosity, to the values measured in other $z>6$ quasars \citep[data from][]{pensabene21,pensabene22} as a function of the upper energy level of the transitions. Strikingly, all of the detected line transitions point to a high H$_2$O- or OH-to-IR luminosity ratio, about 3 times higher than what is observed in other quasars at similar redshifts. The modest number of transitions observed in our study makes it impossible to draw any robust conclusions on the origin of this discrepancy. We speculate on two scenarios: 1) The IR luminosity estimate is correct, in which case the excess may be associated with a low contribution of radiative excitation (as reflected by the non-detection of high-J H$_2$O lines) in PJ183+05 compared to other quasars. Because of the high energies of the involved levels, this scenario would favor the presence of shocks \citep[e.g.,][]{vandishoeck21}. Alternatively, 2) the IR luminosity estimate in PJ183+05 is underestimated. The excellent sampling of the dust continuum shown in Fig.~\ref{fig_sed} pins down the rest-frame far-IR side of the dust emission pretty accurately, but leaves room for an additional hot dust component at mid-IR frequencies (such as an AGN torus), where various H$_2$O and OH transitions relevant for the IR pumping mechanism reside.

\subsection{Line profiles and relative velocities}

Because of their relatively high energies, water levels at J$_{\rm up}>2$ and hydroxyl levels above the ground state are expected to be significantly populated via collisions only in the presence of shocks. Furthermore, P-Cygni profiles have been observed in various water and hydroxyl transitions in lower--redshift, IR--luminous galaxies, thus demonstrating that these species trace dense molecular outflows \citep[e.g.,][]{goicoechea05,goicoechea06}. Similarly, high-ionization lines such as optical \Oiii{} commonly show blue-shifted wings and broadened spectral profiles that reveal the presence of outflows of the ionized medium \citep[e.g.,][]{nesvadba08,carniani15,bischetti17}.

\begin{figure}
\begin{center}
\includegraphics[width=0.99\columnwidth]{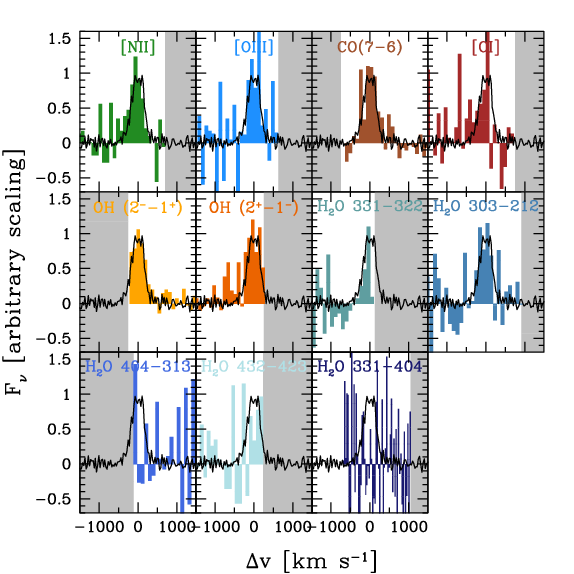}\\
\end{center}
\caption{Comparison between the line profiles of the main transitions in our study (shaded histograms) and the \Cii{} line profile (black histograms). The grey shading masks regions that are not covered within our observations or are contaminated by other emission lines. To first order, all the lines appear to have a consistent bulk velocity and width. A tentative H$_2$O absorption is detected at -750\,\kms{} in the 3(3,1)-3(2,2) and 3(0,3)-2(1,2) transitions, suggesting the presence of an outflow.}
\label{fig_dvel}
\end{figure}

In Fig.~\ref{fig_dvel} we compare the spectral profiles of all the transitions studied in our work. We include in the comparison the three J$_{\rm up}\leq 4$ water transitions that remain undetected in our study. We compare each transition to the \Cii{} line, which has the highest S/N. The velocity profiles of all the detected transitions appear in good agreement at the S/N levels of our observations. All the detected lines show negligible velocity shifts along the line of sight compared to the rest frame of the galaxy as defined by the \Cii{} line, with all the velocity shifts, $\Delta v=c\,\delta z / (1+z)$, significantly smaller than the adopted channel width (90\,\kms{}). The line width is to first order identical in all the transitions. We do not find any clear evidence of P-Cygni profiles. We note the presence of a tentative absorption feature shifted by about -750\,\kms{} in the H$_2$O 3(3,1)-3(2,2), 3(0,3)-2(1,2), and possibly 4(3,2)-4(2,3) lines. Noticeably, \citet{butler22} found a similar absorption feature in the OH 119\,$\mu$m transition for PJ183+05, suggesting that the H$_2$O absorption found here is real. We note that this absorbing component does not match the velocity of the proximate damped-Ly$\alpha$ absorber reported in \citet{banados18b} (which shows a velocity offset of $\approx 1400$\,\kms{}).

\subsection{Masses}\label{sec_masses}

\begin{figure}
\begin{center}
\includegraphics[width=0.99\columnwidth]{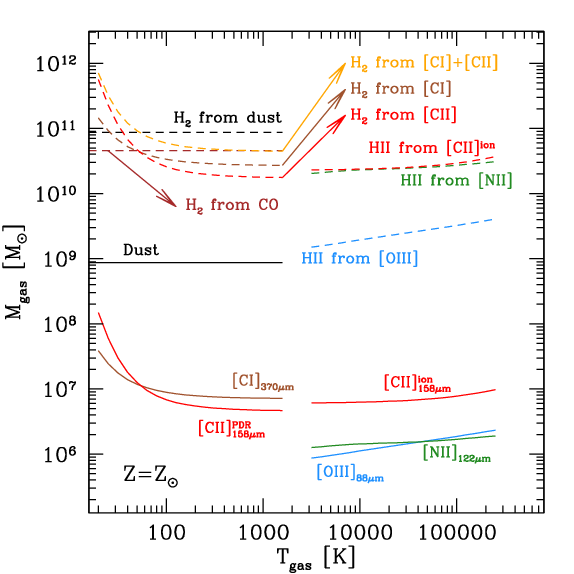}\\
\end{center}
\caption{Mass budget in the ISM of the quasar PJ183+05. Solid lines refer to the masses in dust (black), neutral carbon (brown), singly-ionized carbon (red), singly-ionized nitrogen (green), and doubly-ionized oxygen (blue). The dashed lines show the corresponding molecular and ionized hydrogen mass estimates, after correcting for abundances following \citet{nicholls17} in the case of Z=Z$_\odot$, for a gas-to-dust mass ratio of 100, and for $r_{71}=0.38$ and $\alpha_{\rm CO}=0.8$\,\Msun{}\,(\Kkmspc)$^{-1}$ following \citet{carilli13}. For the ionized medium, we assume a gas density $n=180$\,cm$^{-3}$. We find that the molecular component amounts to $M_{\rm H2}=(0.5-1)\times 10^{11}$\,\Msun{}. The mass of ionized gas traced by low-ionization lines is $(2-3)\times10^{10}$\,\Msun{}, while the highly-ionized gas phase accounts for $(1-5)\times10^{9}$\,\Msun{}.}
\label{fig_masses}
\end{figure}

Having constrained the excitation conditions of the ionized species examined in our work (Fig.~\ref{fig_fsl_ratios}), we can use the information on the relative population of the energy levels to infer the total mass in each ion, and thus the total gas mass via our abundance assumptions. We show these gas estimates in Fig.~\ref{fig_masses}, where we compare them with the mass estimates inferred from the dust in sec.~\ref{sec_continuum}, scaled assuming $\delta_{\rm g/d}=100$ \citep{berta16}; from CO(7-6), assuming $r_{71}=0.38$ and $\alpha_{\rm CO}=0.8$\,\Msun{}\,(\Kkmspc)$^{-1}$ \citep{carilli13}; and from \Ci{} and \Cii$^{\rm PDR}$, following \citet{weiss03,weiss05} and \citet{venemans17b}, where we assume $T_{\rm ex}=T_{\rm gas}$. We find that carbon fine-structure lines suggest the presence of $\sim10^7$\,\Msun{} in both neutral and singly-ionized carbon. On the other hand, from the observed luminosities of \Nii{} and \Oiii{}, we infer $\sim 2\times10^6$\,\Msun{} in each ion. Unsurprisingly, the mass estimates based on \Ci{} and \Cii{}$^{\rm PDR}$ are sensitive to the adopted kinetic energy of the collision partners at $T_{\rm gas}\lsim 100$\,K, i.e., comparable to the $^3$P$_{1,2}$ and $^3$P$_{3/2}$ energy levels, respectively. 

We scale these mass estimates by the abundances computed following eqs.~\ref{eq_C_O}--\ref{eq_N_O} for Z=Z$_\odot$ \citep{nicholls17}. We find that the low-ionization species (\Cii{}, \Nii{}) yield very consistent ionized gas mass estimates, $M_{\rm HII}=(2-3)\times 10^{10}$\,\Msun{}. The mass estimate based on \Oiii{} is $\gsim 5\times$ lower, likely due to the fact that only a small fraction of oxygen atoms are highly ionized. As for the molecular medium, the mass estimates based on singly-ionized and neutral carbon are comparable. We sum them to infer the total mass associated with carbon atoms and ions. We find a corresponding molecular gas mass $M_{\rm H2, [CI]+[CII]}\approx 5\times10^{10}$\,\Msun{} at $T_{\rm gas}\gsim 100$\,K. This is in excellent agreement with the CO-based mass estimate, $M_{\rm H2, CO}=4.5\times10^{10}$\,\Msun{}, and a factor $\sim 2$ lower than the dust-based estimate, $M_{\rm H2, dust}=8.7\times10^{10}$\,\Msun{}. 

While all the mass estimates rely on uncertain conversion factors \citep[see discussions in, e.g.,][]{dunne21,dunne22,decarli22}, all the methods independently converge towards a massive gaseous reservoir in both molecular [$(0.5-1)\times10^{11}$\,\Msun{}] and ionized [$(2-3)\times 10^{10}$\,\Msun{}] phase.

\subsection{Star formation rate estimates}\label{sec_sfr}

Assuming that star formation is the main driver of the gas excitation and photoionization, and of the dust heating, we can convert some of the observed line and continuum luminosities into SFR estimates, using empirical scaling relations. In particular, following the scaling values reported in Tab.~2 by \citet{delooze14}, we find SFR$_{\rm [OIII]}$=110\,\Msun{}\,yr$^{-1}$ and SFR$_{\rm [CII]}$=750\,\Msun{}\,yr$^{-1}$. On the other hand, following \citet{kennicutt12}, we estimate a dust-based SFR$_{\rm IR}$=1330\,\Msun{}\,yr$^{-1}$. 

The discrepancies in the SFR estimates are much larger than formal measurement uncertainties. The scatter in the adopted SFR--luminosity relations is $\lsim 0.3$\,dex, insufficient to account for a $\sim 1$\,dex discrepancy between the \Oiii{}- and IR-based SFR estimates. Hence, the differences must have a more fundamental origin. One way to do so is by correcting for opacity as in Sec.~\ref{sec_opacity}. Doing so, we find SFR$_{\rm [OIII]}$=500\,\Msun{}\,yr$^{-1}$ and SFR$_{\rm [CII]}$=1190\,\Msun{}\,yr$^{-1}$. Another way to reconcile the different SFR estimates is to account for the different star-formation time scales probed by \Cii{} and dust ($\sim 100$\,Myr) and doubly-ionized oxygen ($\sim 10$\,Myr; see \citealt{kennicutt12}).  Finally, other mechanisms might be at play. 
For example, if star formation is heavily dust-enshrouded, dust grains may absorb and reprocess the far-UV photons required to doubly-ionize oxygen atoms, thus suppressing \Oiii{} emission. In addition to these caveats, the uncertainties in the relative abundances, metallicity, and excitation properties discussed in the previous subsections should serve as a warning on the applicability of SFR prescriptions, especially when only one or two independent tracers are available.

\section{Conclusions}\label{sec_conclusions}

We presented a multi-line study at FIR wavelengths of the luminous quasar PJ183+05 at $z=6.4386$. We reported the detection of \Cii{}$_{\rm 158\mu m}$, \Oiii{}$_{\rm 88\mu m}$, \Nii{}$_{\rm 122\mu m}$, \Ci{}$_{\rm 370\mu m}$, CO(7-6), OH$_{\rm 163\mu m}$, and two H$_2$O transitions, and we place limits on the line emission of various high-J CO lines and other water transitions. At the same time, we sampled the dust continuum over a large fraction of the FIR band. We compared our measurements with \textsf{Cloudy} models and analytical prescriptions, with the goal to infer an insight of the physics of the interstellar medium in the host galaxy of a luminous quasar at cosmic dawn. We find that:
\begin{itemize}
\item[{\em i-}] The dust emission is well described by a modified black body with $T_{\rm dust}=47.0_{-2.0}^{+1.5}$\,K, an emissivity index $\beta=1.84_{-0.16}^{+0.15}$, a dust mass of log\,$M_{\rm dust}$/\Msun{}=$8.94_{-0.05}^{+0.06}$, and a size of the emitting region of $\Omega_{\rm s}=0.155_{-0.022}^{+0.029}$\,arcsec$^2$. This implies a non-negligible dust opacity at the frequency of \Cii{} of $\tau_{\rm 1900\,GHz}=0.48\pm 0.04$, and a total infrared luminosity of log\,$L_{\rm IR}$/\Lsun=$12.98_{-0.03}^{+0.04}$. 
\item[{\em ii-}] The \Nii{}$_{\rm 205\mu m}$ emission, inferred via empirical relations, is log\,$L_{\rm [NII]}$/\Lsun=$8.51\pm0.23$. When compared with the observed \Cii{} luminosity, we find that about half of the \Cii{} emission arises from the ionized medium. 
\item[{\em iii-}] The luminosity ratio between the two \Nii{} far-infrared fine-structure lines suggests that the ionized gas has a density of log\,$n$ [cm$^{-3}$]=$2.26\pm0.35$. Using our model of the energy level population, we use this estimate to infer a fiducial estimate of the volume in H{\sc ii} regions, $\sim 0.2$\,kpc$^3$. Assuming that this volume is organized in a collection of $N^{\rm HII}$ spherical H{\sc ii} regions, this implies a Stromgren radius of $\sim 360\,(N^{\rm HII})^{-1/3}$\,pc.
\item[{\em iv-}] The \Oiii$_{\rm 88\mu m}$/\Nii$_{\rm 122\mu m}$ luminosity ratio places loose constraints on the gas metallicity, Z, but tight constraints on the hardness of the photoionization. \textsf{Cloudy} models suggests that the photoionization source has a spectrum comparable to a black body with $T_*\approx25,000$\,K or a single stellar population observed $\approx 9$\,Myr after the burst (assuming for Z=Z$_\odot$). This implies that B-type stars with a luminosity of $\sim10,000$\,\Lsun{} and a main-sequence lifetime of $\sim10$\,Myr are the main drivers of the photoionization.
\item[{\em v-}] The observed \Cii{}/\Ci{} line ratio, and the non-detection of high-J CO transitions, suggest that the neutral/molecular medium is not dominated by X-ray excitation, although we cannot exclude the presence of an X-ray Dominated Region. 
\item[{\em vi-}] Water and hydroxyl line emission appears brighter than what one would expect based on the IR luminosity. This is indicative of the presence of shocks that collisionally excite the higher energy levels of the H$_2$O and OH molecules, or of a warm dust component that is not apparent from the dust continuum sampling in the far-infrared range presented here.
\item[{\em vii-}] The line profiles of all the line transitions studied in this work are consistent at the level of S/N of our observations, irrespective of the gas phase they arise from (ionized, neutral, or molecular). A tentative absorption feature at $\Delta v\sim -750$\,\kms{} is seen in the spectra of two water transitions. Further support to the robustness of this feature comes from a similar absorption reported in the OH$_{\rm 119\mu m}$ line from \citet{butler22}.
\item[{\em viii-}] The mass estimates in ionized and molecular gas, computed based on different tracers, point to a massive gaseous reservoir in both phases, with $(2-3)\times 10^{10}$\,\Msun{} in ionized gas, and $(0.5-1.0)\times 10^{11}$\,\Msun{} in the molecular phase.
\item[{\em ix-}] The host galaxy of PJ183+05 is forming stars at high rates, SFR$_{\rm IR}=1330$\,\Msun{}\,yr$^{-1}$. Estimates based on either \Cii{} and \Oiii{} point to lower values, SFR$_{\rm [CII]}=750$\,\Msun{}\,yr$^{-1}$ and SFR$_{\rm [OIII]}=110$\,\Msun{}\,yr$^{-1}$. The discrepancy is partially mitigated if we account for the dust opacity suppressing the observed line luminosities. In addition, different time scales probed by \Oiii{}, \Cii{} and dust, as well as other processes (e.g., the `\Cii{} deficit') might be responsible for the SFR discrepancies.
\end{itemize}
This study demonstrates the power of multi-line observations in understanding the dominant radiative processes and astrophysical mechanisms at play in galaxies at cosmic dawn. These studies are possible in particular thanks to ALMA's exceptional sensitivity. For instance, other observations of molecules and atoms in PJ183+05 could further expand our understanding of the physical conditions in the star--forming ISM in this quasar. Expanding multi-line investigations to other sources will enable to test whether other quasars show the properties found in PJ183+05. The {\em James Webb Space Telescope} is poised to further expand this field of research, by complementing the study of the rest-frame FIR wavelengths with key diagnostics based on rest-frame optical tracers of the interstellar medium.

\section*{Acknowledgments}

We are grateful to the anonymous A\&A referee for their constructive feedback on the manuscript.
RD is grateful to D.~Romano, R.~Meyer, K.~Butler and P.~van der Werf for inspiring discussions. AP acknowledges support from Fondazione Cariplo grant no.~2020-090.

Data: 2015.1.01115.S and 2016.1.00226.S. 

ALMA is a partnership of ESO (representing its member states), NSF (USA) and NINS (Japan), together with NRC (Canada), NSC and ASIAA (Taiwan), and KASI (Republic of Korea), in cooperation with the Republic of Chile. The Joint ALMA Observatory is operated by ESO, AUI/NRAO and NAOJ.

\begin{appendix}

\section{Analytical vs.\ \textsf{Cloudy} modeling}\label{sec_model_vs_cloudy}

\begin{figure}
\begin{center}
\includegraphics[width=0.99\columnwidth]{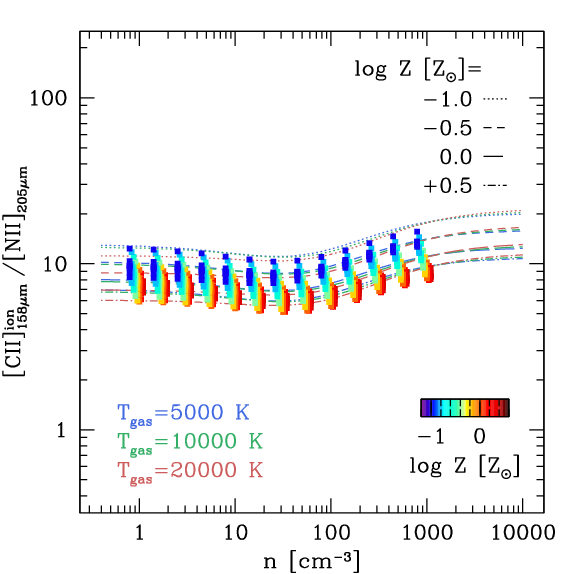}
\end{center}
\vspace{-5mm}
\caption{The luminosity ratio between the \Cii{} luminosity arising from the ionized medium only, \Cii{}$_{158\,\mu{\rm m}}^{\rm ion}$, and the \Nii{}$_{205\,\mu{\rm m}}$ line luminosity, as a function of the ionized gas density $n$. Colored squares mark the results of our \textsf{Cloudy} models, color coded by the gas metallicity. The models were performed on a grid of 0.25 dex spaced in $\log n$, and slightly horizontally displaced in the plot for the sake of clarity. We also show our analytical predictions for the same line ratio, color coded by the gas temperature $T_{\rm gas}$. Dotted, short-dashed, long-dashed, and dot-dashed lines mark models at increasing gas metallicity. The predicted line ratio is unaltered by $T_{\rm gas}$, and is only slightly dependent on $n$, whereas it is linearly dependent on the relative abundance of carbon and nitrogen ions,  [C$^+$/N$^+$]. 
}
\label{fig_cii_nii_predict}
\end{figure}

Fig.~\ref{fig_cii_nii_predict} compares the predicted \Cii{}$_{158\,\mu{\rm m}}$/\Nii{}$_{205\,\mu{\rm m}}$ luminosity ratio for the ionized medium based on the analytical model described in sec.~\ref{sec_model} and on the \textsf{Cloudy} models described in sec.~\ref{sec_cloudy}, for various assumptions on the gas metallicity Z and temperature $T_{\rm gas}$. We show that the ratio does not depend on $T_{\rm gas}$. It depends on the relative abundance of the ions, but the effect is small (a factor $\sim 2$ for a $\sim 30\times$ variation in metallicity). The predictions from the analytical and the \textsf{Cloudy} models are in excellent agreement.

\begin{figure}
\begin{center}
\includegraphics[width=0.49\textwidth]{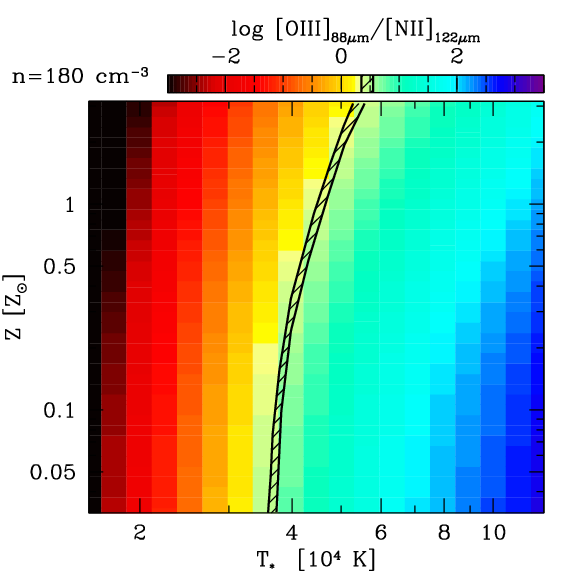}\\
\end{center}
\caption{Same as Fig.~\ref{fig_fsl_ratios}, but computed via the analytical modeling described in sec.~\ref{sec_model}, assuming a gas density of $n=180$\,cm$^{-3}$ and a gas temperature of $T_{\rm gas}=10,000$\,K. The two predictions agree on qualitative terms but not quantitatively, with our simplistic analytical prescriptions underestimating the \Oiii{}$_{\rm 88\,\mu m}$/\Nii{}$_{\rm 122\,\mu m}$ luminosity ratio compared to the superior \textsf{Cloudy} models. 
}
\label{fig_fsl_ratios_model}
\end{figure}

Fig.~\ref{fig_fsl_ratios_model} shows the \Oiii{}$_{88\,\mu{\rm m}}$/\Nii{}$_{122\,\mu{\rm m}}$ luminosity ratio predicted by our analytical model as a function of metallicity Z and of the hardness of the photoionizing radiation field, in the case of a black body with varying effective temperature $T_*$. The prediction is in qualitative agreement with what obtained via \textsf{Cloudy} (see Fig.~\ref{fig_fsl_ratios}), although the analytical model tends to underestimate the predicted line ratio for a given $T_*$. As a consequence, our observations would point to a higher $T_*$ than the one inferred based on the \textsf{Cloudy} models. The effect is however small (a factor $\sim 1.7$ in $T_*$).

\end{appendix}

\label{lastpage}

\end{document}